\documentclass[
aps,%
12pt,%
final,%
notitlepage,%
oneside,%
onecolumn,%
nobibnotes,%
nofootinbib,%
superscriptaddress,%
noshowpacs,%
centertags]%
{revtex4}

\usepackage{graphicx}
\usepackage{dcolumn}
\usepackage{bm}
\usepackage{mathtools}
\usepackage{color}
\usepackage{general_latex}

\begin{document}

\title{Nonclassical properties and quantum resources of hierarchical photonic superposition states}

\author{\firstname{T.J.}~\surname{Volkoff}}
\email{adidasty@gmail.com}
\affiliation{Department of Chemistry, University of California, Berkeley, California 94720, USA}%

\begin{abstract}
We motivate and introduce a class of ``hierarchical'' quantum superposition states of $N$ coupled quantum oscillators. Unlike other well-known multimode photonic Schr\"{o}dinger cat states such as entangled coherent states, the hierarchical superposition states are characterized as two-branch superpositions of tensor products of single-mode Schr\"{o}dinger cat states. In addition to analyzing the photon statistics and quasiprobability distributions of prominent examples of these nonclassical states, we consider their usefulness for high-precision quantum metrology of nonlinear optical Hamiltonians and quantify their mode entanglement. We propose two methods for generating hierarchical superpositions in $N=2$ coupled microwave cavities which exploit currently existing quantum optical technology for generating entanglement between spatially separated electromagnetic field modes.
\end{abstract}
\maketitle

\section{Introduction\label{sec:intro}}
Maximally entangled quantum states occupy a distinguished position in the theory of quantum information. One has only to consider the central role of Greenberger-Horne-Zeilinger (GHZ) states of $(\mathbb{C}^{2})^{\otimes N}$ \cite{ghz} in many quantum algorithms and quantum teleportation protocols \cite{nielsen} to be convinced of their importance.  From a practical perspective, such states comprise the most valuable quantum resource, both in terms of entanglement and usefulness for quantum metrology \cite{pezzesmerz,toth}. Unfortunately, in the case of $N$ two-level quantum systems such as spin-1/2 chains, the maximally entangled states are sensitive to local errors (e.g., phase flips) and can quickly lose all nonclassical resources. However, because of the countably infinite dimension of the Hilbert space of a chain of quantum oscillators (isomorphic to $(\ell^{2}(\mathbb{C}))^{\otimes N}$), one may hope to engineer maximally entangled states of a subspace isomorphic to $(\mathbb{C}^{2})^{\otimes N}$ that are robust under quantum evolutions corresponding to relevant sources of decoherence.

The entangled coherent states (see Ref.\cite{dodonovbook,sandersrev} and references therein) are paradigmatic examples of entangled states of $\ell^{2}(\mathbb{C})\otimes \ell^{2}(\mathbb{C})$. However, only a strict subset of these are maximally entangled. Explicit conditions for maximal entanglement of linear combinations of products of coherent states have been found \cite{najarbashi} and multimode entangled coherent states have been studied in the context of encoding continuous variable quantum information \cite{clusterecs,ralph,milburncats}. To extend the notion of GHZ states to separable Hilbert space (i.e., Hilbert space with countable orthonormal basis) in a general way, we first introduce the following two-branch, $N$ mode states:
\begin{equation}
{1\over \sqrt{2+2\text{Re}(z^{N})}}\left( \mathbb{I}_{N} + U^{\otimes N}\right) \ket{\phi} \, , \, z:= \langle \phi \vert U\vert \phi \rangle 
\label{eqn:ecsgen}\end{equation}

where $\ket{\phi}$ is a single-mode pure quantum state of Hilbert space $\mathcal{H}$, and $U$ is a partial isometry with $\ket{\phi}$ in its domain. When $z=0$, the branches $\ket{\phi}^{\otimes N}$ and $U^{\otimes N}\ket{\phi}^{\otimes N}$ are orthogonal in $\mathcal{H}^{\otimes N}$ and the resulting state is a strict analog of a GHZ state. For example, superpositions of oscillator Fock states proportional to $\ket{n}^{\otimes N}+e^{i\varphi}\ket{m}^{\otimes N}$, with $\varphi \in [0,2\pi]$, fall into this class and are interesting for their quantum optical properties. Such superpositions represent the most obvious generalization of GHZ states to the Hilbert space of $N$ oscillators, $(\ell^{2}(\mathbb{C}))^{\otimes N}$. The case of $z\neq 0$, although deviating from the strict notion of GHZ states due to non-orthogonality of the branches, contains many important macroscopic $N$-mode superpositions. The entangled coherent states (having $\ket{\phi} = \ket{\alpha}$ and $U = e^{i\theta a^{\dagger}a}D(\beta)$, where $\theta \in [0,2\pi]$ and $D(\beta)$ is the oscillator displacement operator for $\alpha$, $\beta \in \mathbb{C}$) serve as well-studied examples. However, the focus of this paper will be on revealing certain new states of the $z=0$ set. In particular, consider taking $\ket{\phi} = \ket{e_{1}}:= \left( {\mathbb{I} + V \over \sqrt{2+2w}} \right)\ket{\phi'}$, $\ket{\phi ' } \in \mathcal{H}$, where $V$ a single-mode partial isometry containing $\ket{\phi '}$ in its domain and $w=\langle \phi ' \vert V \vert \phi '\rangle \in \mathbb{R}$. Then the state $\ket{e_{2}}:= \left( {\mathbb{I} - V \over \sqrt{2-2w}} \right)\ket{\phi'}$ is orthogonal to $\ket{e_{1}}$. Taking $U = e^{i\theta / N}(\ket{e_{1}}\bra{e_{2}} + \ket{e_{2}}\bra{e_{1}})$ produces two-branch superpositions of the following form, which we refer to as hierarchical cat states (HCS):

\begin{equation}
\ket{\text{HCS}_{N}^{\theta}} := {1\over \sqrt{2}}\left(  \left( {(\mathbb{I}+V)\ket{\phi '} \over \sqrt{2+2w}}\right)^{\otimes N} +e^{i\theta} \left( {(\mathbb{I}-V)\ket{\phi '} \over \sqrt{2-2w}} \right)^{\otimes N} \right)
\label{eqn:hcs}\end{equation}

The origin of the name ``hierarchical cat state'' is self-evident: $\ket{\text{HCS}_{N}^{\theta}}$ is an equal weight superposition of two orthogonal branches in $\mathcal{H}^{\otimes N}$ (i.e., an $N$-mode ``cat'' state), while each branch is a tensor product of ``kitten'' superpositions in the single-mode Hilbert space $\mathcal{H}$. This construction allows to consider maximally entangled states of a $(\mathbb{C}^{2})^{\otimes N}$ subspace of $(\ell^{2}(\mathbb{C}))^{\otimes N}$ that retain single-mode quantum coherence in the states $\ket{\phi '} \pm V\ket{\phi '}$ even after intermode decoherence processes reduce the superposition state $\ket{\text{HCS}_{N}^{\theta}}$ to an equal probability statistical mixture of tensor products.

When $\mathcal{H} \cong \ell^{2}(\mathbb{C})$, $\ket{\text{HCS}_{N}^{\theta}}$ is in a two (complex) dimensional subspace of $(\ell^{2}(\mathbb{C}))^{\otimes N}$. By appropriate choices of $V$ and $\ket{\phi '}$, the branches of $\ket{\text{HCS}_{N}^{\theta}}$ can take the form of tensor products of single-mode photonic Schr\"{o}dinger cat states such as the even and odd coherent states $\ket{\psi_{\pm}} \propto \ket{\alpha} \pm \ket{-\alpha}$ \cite{dodonovart} or superpositions of squeezed states.  For these photonic HCS states, it is clear that if the single-mode coherence time is sufficiently long (e.g., greater than the intermode coherence time), a statistical mixture of $N$-mode Schr\"{o}dinger cat states remains even after intermode coherence is lost by some decohering process. 

The photonic HCS state obtained by taking $V=e^{i\pi a^{\dagger}a}$ to be the oscillator $\pi$ phase shift,  $\ket{\phi '}=\ket{\alpha}$ a coherent state of the quantum oscillator with amplitude $\vert \alpha \vert$, and $\theta = 0$ or $\theta = \pi$ in (\ref{eqn:hcs}) was introduced in Ref.\cite{volkoff}. We will label these states by $\ket{\text{HCS}_{N}^{\pm}(\alpha)}$ (the $+$ symbolizing $\theta = 0$ and the $-$ symbolizing $\theta = \pi$) and they will serve as the canonical examples of photonic HCS in this work. Each branch of $\ket{\text{HCS}_{N}^{\pm}(\alpha)}$ is an $N$-fold tensor product of either the even coherent state $\ket{\psi_{+}}$ or odd coherent state $\ket{\psi_{-}}$. If $N$ is an odd natural number, then the even (odd) branch is an eigenvector of the photon parity operator $e^{i\pi \sum_{j=1}^{N}a^{\dagger}_{j}a_{j}}$ with eigenvalue  $ 1$ ($-1$) and hence, $e^{i\pi \sum_{j=1}^{N}a^{\dagger}_{j}a_{j}}\ket{\text{HCS}_{N}^{\pm}(\alpha)} = \ket{\text{HCS}_{N}^{\mp}(\alpha)}$. If $N$ is an even natural number, $\ket{\text{HCS}_{N}^{\pm}(\alpha)}$ is invariant under such a local $\pi$ rotation. Independent of $N$, a photon parity measurement results in a projection of $\ket{\text{HCS}_{N}^{\pm}(\alpha)}$ onto either the even or odd branch. The entangled coherent states $\ket{\text{ECS}_{N}^{\pm}(\alpha)}\propto \ket{\alpha}^{\otimes N} \pm \ket{-\alpha}^{\otimes N}$ are invariant under the bosonic algebra freely generated by $a_{i}a_{j}$, for $i$, $j\in \lbrace 1,\ldots N\rbrace$, so that they are considered to be Barut-Girardello coherent states of $\mathfrak{sp}(N,\mathbb{C})$ \cite{trifonov}. The state $\ket{\text{HCS}_{N}^{\pm}(\alpha)}$ is invariant under the algebra freely generated by the identity operator and the two-photon annihilation operators $a_{j}^{2}$, $j=1,\ldots ,N$. This algebraic property allows $\ket{\text{HCS}_{N}^{\pm}(\alpha)}$ to be considered as superpositions of $\mathfrak{sp}(N,\mathbb{C})$ Barut-Girardello coherent states. However, while such properties as quasiprobability densities, photon statistics, and dissipative evolutions of the entangled coherent states have been thoroughly documented \cite{manko,dodonovlaser}, a detailed analysis of the properties of photonic HCS states is lacking.

When $\ket{\text{HCS}_{2}^{\pm}(\alpha)}$ is shared between two spatially separated parties, the state $\ket{\text{HCS}_{2}^{\pm}(\alpha)}$ serves as an entanglement resource for teleportation of an arbitrary superposition of coherent states of the form $ c_{1}\ket{\alpha} +c_{2}\ket{-\alpha}$ in the same way that the GHZ state $1/\sqrt{2}(\ket{0}^{\otimes N} +\ket{1}^{\otimes N})$ serves as an entanglement resource for teleportation of an arbitrary qubit pure state. This follows from the fact that $\ket{\text{HCS}_{2}^{\pm}(\alpha)}$ are maximally entangled states in the 4-dimensional sub-Hilbert space spanned by $\lbrace \ket{(-1)^{j}\alpha}\otimes\ket{(-1)^{\ell}\alpha} \rbrace_{j, \ell \in \lbrace 0, 1 \rbrace}$ \footnote{In fact, the states $\ket{\text{HCS}_{2}^{\pm}(\alpha)}$ exhibit the same amount (1 ebit) of entanglement entropy as $\ket{\text{ECS}^{-}_{2}(\alpha)}$ \cite{vanenk,wang}. A ``Bell basis'' of maximally entangled states for this 4-dimensional sub-Hilbert space is given by: $\ket{\text{HCS}_{2}^{+}(\alpha)}$, $\ket{\text{HCS}_{2}^{-}(\alpha)}$, $\ket{\text{ECS}_{2}^{-}(\alpha)}$, $(e^{i\pi a^{\dagger}a}\otimes \mathbb{I})\ket{\text{ECS}_{2}^{-}(\alpha)}$ \cite{jeongbell}.}. The states $\ket{\text{HCS}_{N}^{\pm}(\alpha)}$ are also useful probes for high-precision phase estimation of Hamiltonians of $\mathfrak{sl}(2,\mathbb{C})$ \cite{volkoff4} (see also Section \ref{sec:metrousesec}). These intriguing attributes motivate a more thorough description and analysis of HCS states.

The remainder of this paper is structured as follows: in Section \ref{sec:nonclass}, we indicate some nonclassical properties of $\ket{\text{HCS}_{N}^{\pm}(\alpha)}$, citing the nonclassical properties of the entangled coherent states for comparison; Section III is devoted to exploration of the quantum resources, in particular the metrological usefulness and entanglement entropy, of $\ket{\text{HCS}_{N}^{\pm}(\alpha)}$; in Section IV, we propose two methods for generating $\ket{\text{HCS}_{N}^{+}(\alpha)}$ using techniques which are accessible by current quantum optical technology; in Section V, we demonstrate how the idea of hierarchically encoding  continuous variable quantum information can be deepened with many levels of hierarchy. We do not attempt an exhaustive analysis of photonic hierarhical cat states, but rather try to show the most salient properties of these states by considering basic examples.

\section{Nonclassical properties of HCS\label{sec:nonclass}}
Here, we take note of the basic photon statistics of $\ket{\text{HCS}_{N}^{+}(\alpha)}$ for arbitrary $N$, derive some of its quasiprobability distributions for $N=2$, and show a duality between Pauli matrices and photon operations in the subspace of $\ell^{2}(\mathbb{C})$ spanned by the even and odd coherent states. Throughout this section, we compare the nonclassical properties of $\ket{\text{HCS}_{N}^{\pm}(\alpha)}$ to those of the entangled coherent states, which are more familiar. We show the inner products of $\ket{\text{ECS}_{N}^{\pm}(\alpha)}$ and $\ket{\text{HCS}_{N}^{\pm}(\alpha)}$ immediately:
\begin{equation}
\langle \text{ECS}_{N}^{+}(\alpha) \vert \text{HCS}_{N}^{+}(\alpha) \rangle = {1\over \sqrt{1+e^{-2N\alpha^{2}}}} \left\{
  \begin{array}{lr}
    ({1\over 2} + {1\over 2}e^{-2\alpha^{2}})^{N/2} + ({1\over 2} - {1\over 2}e^{-2\alpha^{2}})^{N/2} & , N \text{ even} \\
    ({1\over 2} + {1\over 2}e^{-2\alpha^{2}})^{N/2}& , N \text{ odd}
  \end{array}
\right.
\end{equation}
\begin{equation}
\langle \text{ECS}_{N}^{+}(\alpha) \vert \text{HCS}_{N}^{-}(\alpha) \rangle = {1\over \sqrt{1+e^{-2N\alpha^{2}}}} \left\{
  \begin{array}{lr}
    ({1\over 2} + {1\over 2}e^{-2\alpha^{2}})^{N/2}& , N \text{ even} \\
    ({1\over 2} + {1\over 2}e^{-2\alpha^{2}})^{N/2} - ({1\over 2} - {1\over 2}e^{-2\alpha^{2}})^{N/2} & , N \text{ odd}
  \end{array}
\right.
\end{equation}
\begin{equation}
\langle \text{ECS}_{N}^{-}(\alpha) \vert \text{HCS}_{N}^{\pm}(\alpha) \rangle =  \left\{
  \begin{array}{lr}
    0& , N \text{ even} \\
\pm    { ({1\over 2}-{1\over 2}e^{-2\alpha^{2}})^{N/2} \over \sqrt{1-e^{-2N\alpha^{2}}}} & , N \text{ odd}
  \end{array}
\right.
\end{equation}
where we have taken $\alpha \in \mathbb{R}$. It is intriguing to take note of the $\alpha \rightarrow \infty$ asymptotics of these inner products (the $N\rightarrow \infty$ asymptotic is always zero). For any odd $N$ and for reasonably large $\alpha$, $\langle \text{ECS}_{N}^{+}(\alpha) \vert \text{HCS}_{N}^{-}(\alpha) \rangle \approx 0$, while for any even $N$ and any $\alpha$, $\langle \text{ECS}_{N}^{-}(\alpha) \vert \text{HCS}_{N}^{\pm}(\alpha) \rangle = 0$ identically. The total expected photon number in all of these states is asymptotically $N\alpha^{2}$, i.e., $\ket{\text{HCS}_{N}^{\pm}(\alpha)}$ and $\ket{\text{ECS}_{N}^{\pm}(\alpha)}$ differ mainly in photon statistics and quantum correlations, not in intensity. In particular, if $P_{e}$ ($P_{o} = \mathbb{I}-P_{e}$) is the projection onto the even (odd) photon number subspace\footnote{Explicitly, $P_{e} = \sum_{\pmb{\vec{n}}: \Vert \pmb{\vec{n}} \Vert^{2} = 0\text{ mod}2}P_{\pmb{\vec{n}}}$, where $P_{\pmb{\vec{n}}}$ is the rank one projector onto the ray $\ket{n_{1}}\otimes \cdots \otimes \ket{n_{N}}$.} of $(\ell^{2}(\mathbb{C}))^{\otimes N}$, it is clear that $P_{o}\ket{\text{ECS}_{N}^{+}(\alpha)}=P_{e}\ket{\text{ECS}_{N}^{-}(\alpha)}=0$ while $P_{o}\ket{\text{HCS}_{N}^{+}(\alpha)} = P_{e}\ket{\text{HCS}_{N}^{+}(\alpha)} = 1/\sqrt{2}$.

Because of its symmetry, $\ket{\text{HCS}_{N}^{\pm}(\alpha)}$ has a simple expression as a superposition of tensor products of coherent states. To do this, first consider the direct product $\mathbb{Z}_{2}\times \cdots \times \mathbb{Z}_{2}$ ($N$ times) with group operation addition modulo 2. This group is isomorphic to the abelian group $\mathfrak{U}$ with elements given by $\bigotimes_{j=1}^{N}e^{ik_{j}\pi a^{\dagger}_{j}a_{j}}$, $k_{j} \in \lbrace 0,1\rbrace$, and group operation being operator multiplication. Let $\mathfrak{U}_{1}$ ($\mathfrak{U}_{2}$) be the subgroup of elements corresponding to $\pmb{\vec{k}}$ such that the number of nonzero entries of $\pmb{\vec{k}}$ is even (odd). Then, one can write (again for $\alpha \in \mathbb{R}$):
\begin{eqnarray}
\ket{\text{HCS}_{N}^{\pm}(\alpha)} &=& {e^{N\alpha^{2}/2}\over 2^{N+{1\over 2}}}\left( \sum_{j=1,2} \left( \cosh^{-N/2}\alpha^{2} + (-1)^{j\mp 1}\sinh^{-N/2}\alpha^{2} \right)\sum_{u\in \mathfrak{U}_{j}}u \right) \ket{\alpha}^{\otimes N}.
\end{eqnarray}
From the above expression, the expansion in the Fock state basis can be made explicit by making use of the fact that $\ket{\alpha} = e^{-\vert \alpha \vert^{2}/2}\sum_{n=0}^{\infty}{\alpha^{n}\over \sqrt{n !}}\ket{n}$.

\begin{figure}[t!]
\includegraphics[scale=0.7]{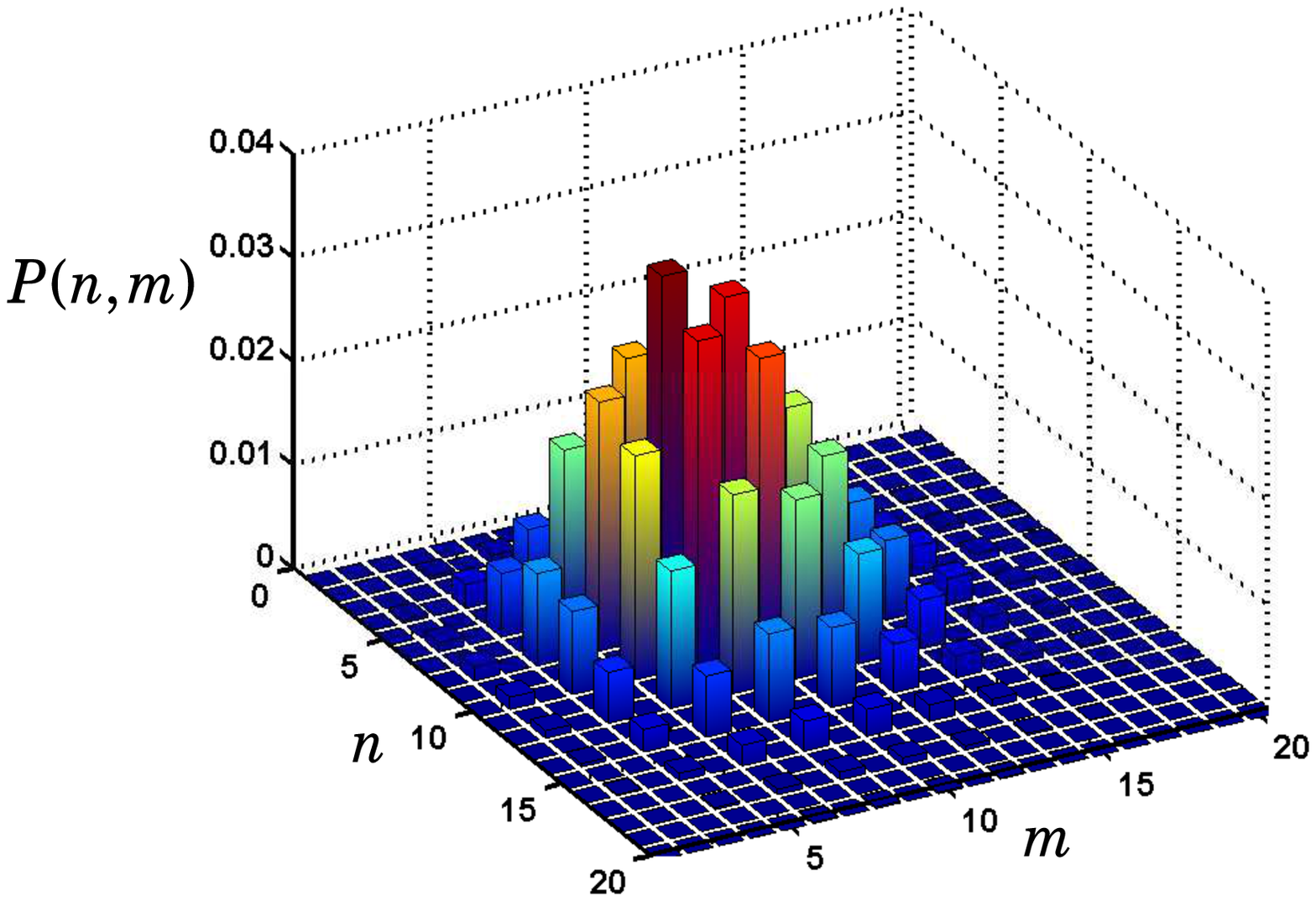}
\caption{Photon number distribution $P(n,m)$ for $\ket{\text{HCS}_{N=2}^{+}(\alpha=3)}$; $n$, $m\in \lbrace 1,\ldots , 20 \rbrace$.\label{fig:pnd}}
\end{figure}

The photon number distribution of $\ket{\text{HCS}_{N}^{+}(\alpha)}$ for any $\alpha \in \mathbb{C}$ is given by:
\begin{eqnarray}
P(\pmb{\vec{n}}) &=& \vert \langle \pmb{\vec{n}} \vert  \text{HCS}_{N}^{+}(\alpha) \rangle \vert^{2} \nonumber 
\\ &=& {e^{-N\vert \alpha \vert^{2}}\over 2^{N+1}} \Bigg\vert \sum_{j=0,1}{1\over (1+(-1)^{j}e^{-2\vert \alpha \vert^{2}})^{N/2}}\prod_{k=1}^{N}(1+(-1)^{n_{k}+j}){\alpha^{n_{k}}\over \sqrt{n_{k}!}}\Bigg\vert^{2}
\end{eqnarray}
where $\ket{\pmb{\vec{n}}}:= \ket{n_{1}}\otimes \cdots \otimes \ket{n_{N}}$ for $\vec{n} \in \mathbb{N}^{N}$. It is clear from the above expression that if $\pmb{\vec{n}}$ does not have all entries even or odd, then the photon number distribution vanishes. For $N=2$, this results in a checkerboard pattern of zero and nonzero probabilities on the lattice $\mathbb{Z}_{\ge 0}\times \mathbb{Z}_{\ge 0}$ (Fig.\ref{fig:pnd}). This feature stands in contrast to the distribution $\vert \langle \pmb{\vec{n}} \ket{\text{ECS}^{+}_{N}(\alpha)}\vert^{2}$, which is identically zero if and only if $\sum_{k=1}^{N}n_{k}$ is odd. Our present focus on the photon statistics of $\ket{\text{HCS}_{N}^{+}(\alpha)}$ is merely due to the fact that they are the hierarchical cat states of most immediate practical use for continuous variable quantum information processing. Indeed, more complex photon statistics are furnished by hierarchical cat states formed from, e.g., the $\mathbb{Z}/4\mathbb{Z}$ coherent states, which generalize the even/odd coherent states by forming a $\mathbb{C}^{4}$ subspace of $\ell^{2}(\mathbb{C})$ having orthonormal basis

\begin{eqnarray}
\ket{e_{0}} &=& {\ket{\alpha} + \ket{-\alpha} + \ket{i\alpha} + \ket{-i\alpha} \over 2e^{-\vert \alpha \vert^{2}/2}\sqrt{2\cosh \vert \alpha \vert^{2} + 2\cos \vert \alpha \vert^{2}} } \nonumber \\
\ket{e_{1}} &=& {\ket{\alpha} - \ket{-\alpha} - i \ket{i\alpha} +i \ket{-i\alpha} \over 2e^{-\vert \alpha \vert^{2}/2}\sqrt{2\sinh \vert \alpha \vert^{2} + 2\sin \vert \alpha \vert^{2}} } \nonumber \\
\ket{e_{2}} &=& {\ket{\alpha} + \ket{-\alpha} - \ket{i\alpha} - \ket{-i\alpha} \over 2e^{-\vert \alpha \vert^{2}/2}\sqrt{2\cosh \vert \alpha \vert^{2} - 2\cos \vert \alpha \vert^{2}} } \nonumber\\
\ket{e_{3}}& =& {\ket{\alpha} - \ket{-\alpha} + i \ket{i\alpha} -i \ket{-i\alpha} \over 2e^{-\vert \alpha \vert^{2}/2}\sqrt{2\sinh \vert \alpha \vert^{2} - 2\sin \vert \alpha \vert^{2}} }
\label{eqn:z4}
\end{eqnarray}
where $\langle n \vert e_{j} \rangle \neq 0$ if and only if $n\equiv j\mod 4$.

\begin{figure}[t!]
\includegraphics[scale=0.5]{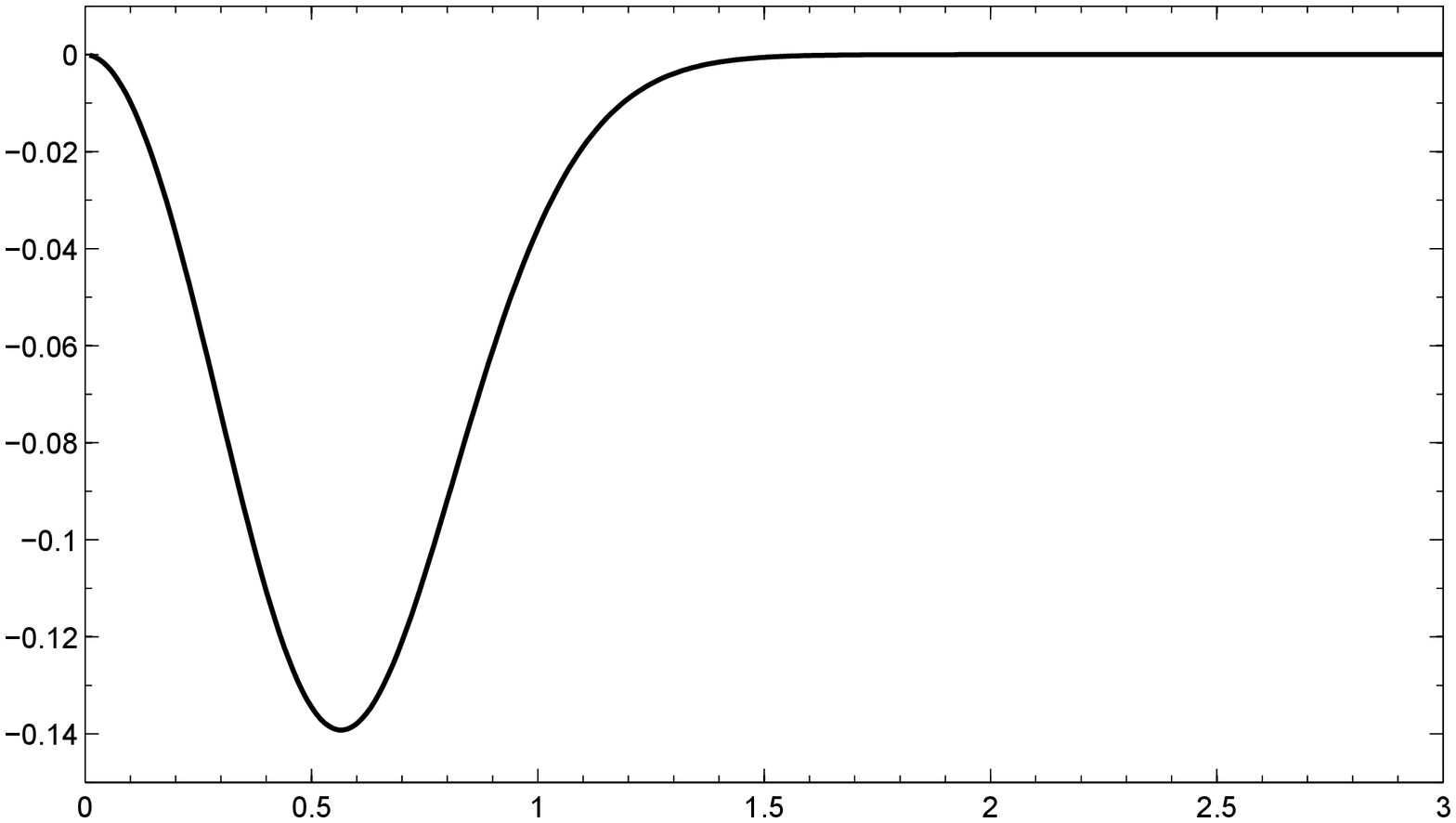}
\caption{Mandel $Q_{M}$ parameter for $\ket{\text{HCS}_{N}^{\pm}(\alpha)}$ for $\vert \alpha \vert \in [0,3]$. \label{fig:mandel}}
\end{figure}

The variance of a single mode quadrature $x^{(\theta)}_{j}:= {1\over \sqrt{2}}(a_{j}e^{-i\theta} + a^{\dagger}_{j}e^{i\theta})$ in $\ket{\text{HCS}_{N}^{\pm}(\alpha)}$ is \begin{equation}
{1\over 2}+  \vert \alpha \vert^{2}(\coth 2\vert \alpha \vert^{2} +\cos(2\text{Arg}\alpha - 2\theta) ) 
\label{eqn:hcssqueeze}
\end{equation} for all $\theta$, i.e., since this variance is greater than $1/2$, $\ket{\text{HCS}_{N}^{\pm}(\alpha)}$ is not squeezed in any quadrature. $\ket{\text{HCS}_{N}^{\pm}(\alpha)}$ do not exhibit second order squeezing \cite{buzek2} (also referred to as amplitude squared squeezing) due to their being eigenvectors of $a_{j}^{2}$ for all $j = 1,\ldots ,N$.  The single mode Mandel $Q_{M}$ parameter of $\ket{\text{HCS}_{N}^{\pm}(\alpha)}$ is always negative, but exhibits a dip for $\vert \alpha \vert \approx 3/2$ as shown in Fig.\ref{fig:mandel}).
We now turn to the quasiprobability distributions and functional representations of $\ket{\text{HCS}_{2}^{\pm}(\alpha)}$ with the focus being on the functional form of the hierarchical coherences. In addition, we utilize the explicit expressions of the quasiprobability distributions to infer nonclassical features of the states.  Each quasiprobability distribution for a quantum state $\rho$ of $(\ell^{2}(\mathbb{C}))^{\otimes N}$ is obtained by Fourier transformation on $\mathbb{C}^{N}$ of a quantum characteristic function corresponding to a particular ordering of bosonic operators \cite{glauber}. It is well known that negative values of the singular quasidistribution (i.e., Sudarshan-Glauber function) for a given quantum state $\rho$ indicate nonclassical photon statistics, i.e., indicate that the photon number distribution is not Poissonian \cite{mandel}. In a similar spirit, the existence of negative values of the Wigner function for a given state indicates non-Gaussian quadrature correlations.  The explicit form for the Wigner function of $\ket{\text{HCS}_{N}^{+}(\alpha)}$ was shown in Ref.\cite{volkoff}, but the analytic expression is only useful for technical purposes. The important point is that the Wigner function of a photonic state has an interpretation as a continuous set of interference experiments. This is clear from the definition of the single mode Wigner function as $W(\gamma)=(2/\pi^{2})\langle D(\gamma)e^{i\pi a^{\dagger}a}D(-\gamma)\rangle$; it shows how a quantum state changes when its coherent state components are displaced by $-\gamma$ in phase space, then are reflected in phase space, and are displaced again opposite to $-\gamma$. Coherence between coherent state components appears as fringes in the Wigner function. Accordingly, the states $\ket{\text{HCS}_{2}^{\pm}(\alpha)}$ exhibit two sources of interference: 1) from the coherence in each branch $\ket{\psi_{\pm}}^{\otimes 2}$, and 2) from the coherence between the branches. These two sources of coherence are not immediately visible from the expression of $\ket{\text{HCS}_{2}^{\pm}(\alpha)}$ as an unequal superposition of tensor products of coherent states.

The existence of zeroes of the $Q$-function of a given quantum state indicates that the singular quasidistribution for this state takes negative values and, hence, exhibits nonclassical features. The $Q$-function of $\ket{\text{HCS}_{N}^{+}(\alpha)}$, which is a true probability distribution on $\mathbb{C}^{N}$, is given by $Q(\ket{\text{HCS}_{N}^{+}(\alpha)})={1\over \pi}\vert \left( \otimes_{k=1}^{N}\ket{\beta_{k}},\ket{\text{HCS}_{N}^{+}(\alpha)} \right) \vert^{2}$. For $N=2$ and $\alpha \in \mathbb{C}$ we have:
\begin{eqnarray}
Q_{\ket{\text{HCS}_{2}^{+}(\alpha)}}(\beta_{1},\beta_{2}) &=& {\pi\over 2}\left( Q_{+}(\beta_{1})Q_{+}(\beta_{2}) + Q_{-}(\beta_{1})Q_{-}(\beta_{2})\right) \nonumber \\ &+& {e^{-(\vert \beta_{1} \vert^{2}+\vert \beta_{2} \vert^{2})}\over 2\pi \sinh (2\vert \alpha \vert^{2})}\left( \sinh (2\text{Re}(\beta_{1}\alpha)) \sinh (2\text{Re}(\beta_{2}\alpha)) \right. \nonumber \\ &-& \left. \sin (2\text{Im}(\beta_{1}\alpha)) \sin (2\text{Im}(\beta_{2}\alpha))  \right).
\end{eqnarray} 
where $Q_{+(-)}(z)$ is the $Q$-function of the even (odd) coherent state \cite{manko}. The $Q$-function of $\ket{\text{HCS}_{2}^{+}(\alpha)}$ vanishes if and only if each of the terms vanishes. Let us take $\beta_{1}$, $\beta_{2}\in \mathbb{C}$ such that $\text{Re}(\beta_{1}\alpha) = \text{Re}(\beta_{2}\alpha)=0$. In addition, we require that $\beta_{1}$, $\beta_{2}$ satisfy: 1) $\text{Im}(\beta_{1}\alpha)={(2k+1)\pi \over 2}$, where $k \in \mathbb{Z}$, 
and 2) $\text{Im}(\beta_{2}\alpha) = m \pi$, where $m\in \mathbb{Z}$. Under these constraints, $Q_{+}(\beta_{1})=Q_{-}(\beta_{2})=0$. For these values, $\sinh (2\text{Re}(\beta_{1}\alpha))\sinh (2\text{Re}(\beta_{2}\alpha)) - \sin (2\text{Im}(\beta_{1}\alpha)) \sin (2\text{Im}(\beta_{2}\alpha)) =0$, and hence the $Q$-function is zero at these points of $\mathbb{C}\times \mathbb{C}$.

In addition to the quasiprobability distributions, the quadrature distribution, calculated as the square modulus of the Schr\"{o}dinger wavefunction, is an especially useful true probability distribution for systems of oscillators. However, the quadrature distribution is quite specific; all that is needed is a representation of the pure state $\ket{\text{HCS}_{2}^{\pm}(\alpha)}$ in a functional Hilbert space. We choose the Bargmann representation \cite{bargmann} because of the fact that such relevant quantities as the Schr\"{o}dinger wavefunction and the singular quasidistribution of a pure state can be derived from the Bargmann representation by the use of variants of the Segal-Bargmann transformation. As an analytic function $f_{\ket{\text{HCS}_{2}^{\pm}(\alpha)}}$ on $\mathbb{C} \times \mathbb{C}$, the state $\ket{\text{HCS}_{2}^{\pm}(\alpha)}$  is represented by
\begin{equation}
f_{\ket{\text{HCS}_{2}^{\pm}(\alpha)}}(z,w)=\sqrt{2}e^{-\vert \alpha \vert^{2}}\left(   {\cosh(\alpha(z+w)) \mp e^{-2\vert \alpha \vert^{2}}\cosh(\alpha(z-w))\over 1-e^{-4\vert \alpha \vert^{2}}} \right).
\end{equation} 
That $f_{\ket{\text{HCS}_{2}^{\pm}(\alpha)}}(z,w)$ takes the form of a sum of unequally weighted functions is a consequence of the fact that $\ket{\text{HCS}_{2}^{\pm}(\alpha)}$ is an unequally weighted superposition of tensor products of coherent states.

The subspace $\mathcal{K} \cong \mathbb{C}^{2}$ of $\ell^{2}(\mathbb{C})$ spanned by the even and odd coherent states $\ket{\psi_{\pm}}$ has the property that certain photonic operators carry out equivalent operations as the Pauli matrices in this subspace. This allows for quantum operations of a two-level system to be interpreted as photonic operations compressed to this subspace. For example, keeping in mind the action of $\sigma_{x}=\ket{\psi_{+}}\bra{\psi_{-}} + \ket{\psi_{-}}\bra{\psi_{+}}$ in $\mathcal{K}$ we note that $a\ket{\psi_{+}} = \alpha \sqrt{\tanh \vert \alpha \vert^{2}} \ket{\psi_{-}} = \alpha \sqrt{\tanh \vert \alpha \vert^{2}} \sigma_{x}\ket{\psi_{+}}$. Considering the Pauli matrices as observables of a spin-1/2 particle, we find the following expressions in terms of self-adjoint photonic observables:
\begin{eqnarray}
\sigma_{x} &=&  {e^{-\vert \alpha \vert^{2}}\sqrt{\sinh(2\vert \alpha \vert^{2})}\over \vert \alpha \vert}P_{\mathcal{K}}x^{(\text{Arg}\alpha)}P_{\mathcal{K}}\nonumber \\
\sigma_{y} &=& {e^{\vert \alpha \vert^{2}}\sqrt{\sinh(2\vert \alpha \vert^{2})}\over \vert \alpha \vert} P_{\mathcal{K}}x^{({\pi \over 2} + \text{Arg}\alpha )}P_{\mathcal{K}} \nonumber \\
\sigma_{z} &=& {1\over 2\text{Re}( \alpha ^{2})}P_{\mathcal{K}}(e^{i\pi a^{\dagger}a}a^{2} + a^{\dagger 2}e^{-i\pi a^{\dagger}a} )P_{\mathcal{K}}
\label{eqn:pauli}\end{eqnarray}
where $P_{\mathcal{K}}$ is the projection to $\mathcal{K}$. These expressions for Pauli matrices show a duality between quadratures and ``magnetization'' in the subspace $\mathcal{K}$. For example, if $\vert \alpha \vert$ is sufficiently large, we have $P_{\mathcal{K}}x^{(\text{Arg}\alpha)}P_{\mathcal{K}} = \sqrt{2}\vert \alpha \vert \sigma_{x} $, i.e., the interaction picture dynamics of a quantum oscillator (restricted to $\mathcal{K}$) driven with unit amplitude is equivalent to a spin-1/2 particle with magnetic moment $\sqrt{2}\vert \alpha \vert$ in a unit magnetic field along the $x$-axis. The Pauli matrices in (\ref{eqn:pauli}) do not have unique expressions in terms of products of photonic operators and $P_{\mathcal{K}}$, due to the fact that one can rewrite $P_{\mathcal{K}}$ as $(1/2\vert \alpha \vert^{2})P_{\mathcal{K}}(a^{2}e^{-2i\text{Arg}\alpha} + a^{\dagger 2}e^{2i\text{Arg}\alpha} ) P_{\mathcal{K}}$. Some of these alternative expressions can be instructive; for example, one can rewrite $\sigma_{z} = P_{\mathcal{K}}\cos(\pi a^{\dagger}a)P_{\mathcal{K}}$. A similar duality can be derived for $\mathfrak{su}(4)$ observables in terms of projectors in the subspace spanned by the list (\ref{eqn:z4}) and photonic operations.

\section{Quantum resources of HCS\label{sec:metrouse}}
\subsection{Metrological usefulness\label{sec:metrousesec}}
We begin this section by recalling the main problem of quantum metrology and how certain quantum states can be utilized for estimation of dynamical parameters at higher precision than any classical states. Given a smooth manifold $M$, let quantum states be encoded by a differentiable map specified by $\lambda \mapsto \rho_{\lambda}$ for $\lambda \in M$. The goal is to estimate the parameter $\lambda$ with greatest possible precision by making use of an optimal quantum measurement and optimal classical post-processing of the measurement results.

In the present section, we are concerned with the special case of estimation of a displacement parameter. In this case, the parameter manifold is a line with real coordinate $x \in \mathbb{R}$ and the state $\rho_{x} := e^{-ix H}\rho_{0}e^{ix H}$ lies on a path parametrized by $x$ and generated by the self-adjoint, $x$-independent operator $H$. If $\lbrace M(dx) \rbrace$ is a positive operator-valued measure (we will refer to $\lbrace M(dx) \rbrace$ as a ``quantum measurement'' or, simply, ``measurement'' from now on) which is an unbiased estimator of $x$, i.e., $x = \int_{\mathbb{R}}x'\text{tr}(\rho_{x}M(dx'))$, then the quantum Cram\'{e}r-Rao theorem \cite{helstrombook,holevo} states

\begin{equation}
\langle (\delta \hat{x}_{M})^{2}\rangle \ge {1 \over \text{tr}(\rho_{0}L^{2})}
\label{eqn:qcr}
\end{equation}

where $\langle (\delta \hat{x}_{M})^{2} \rangle :=  \int_{\mathbb{R}}(x'-x)^{2}\text{tr}(\rho_{x}M(dx'))$ is a general expression for the variance of the quantum measurement and where $L=L^{\dagger}$ is the symmetric logarithmic derivative operator defined by the equation ${d\rho \over dx} = -i[H,\rho] = {1\over 2}(L\rho+\rho L)$. The quantity $\text{tr}(\rho_{0}L^{2})$ is called the quantum Fisher information of $\rho_{0}$ and is constant on the unitary path generated by $H$ \cite{mgaparis}. If $\rho_{0} = \ket{\psi_{0}}\bra{\psi_{0}}$ is pure, then $\text{tr}(\rho_{0}L^{2}) = 4(\text{tr}(\rho_{0}H^{2}) - \text{tr}(\rho_{0}H)^{2})$. Hence, if an experimenter has unconstrained access to measurements saturating inequality (\ref{eqn:qcr}), then a quantum state with larger quantum Fisher information with respect to $H$ can be considered as a more useful resource for estimating the displacement parameter $x$. In this section, we focus on using certain multimode pure states $\ket{\psi_{0}} \in \mathcal{H}^{\otimes N}$ as probes for displacement metrology for paths generated by 1-local Hamiltonians , i.e., $H$ having the form $H= \sum_{j=1}H^{(j)}\otimes \mathbb{I}_{N-1}$. Specifically, the metrological problem at hand consists of: 1) preparation of $N$ oscillator modes in the probe state $\ket{\text{ECS}_{N}^{+}(\alpha)}$ or $\ket{\text{HCS}_{N}^{+}(\alpha)}$, 2) application of a global unitary operator $\otimes_{j=1}^{N}e^{ixH_{j}}$ with $H_{j} = H$ an oscillator Hamiltonian and $x\in \mathbb{R}$, and 3) estimation of $x$ by an optimal separable measurement on the $N$ modes. It is important to note that determination of the optimal separable measurement corresponding to the probe state and Hamiltonian $H$ requires methods of quantum estimation theory; in particular, the optimal measurement does not necessarily correspond with traditional methods of oscillator signal detection such has homodyne detection. In this section, we assume that the optimal measurement can be performed for any $H$ and we determine the set of $H$ for which $\ket{\text{ECS}_{N}^{+}(\alpha)}$ and $\ket{\text{HCS}_{N}^{+}(\alpha)}$ allow for a greater precision in the determination of $x$ than the tensor product branch states $\ket{\pm \alpha}^{\otimes N}$ and $\ket{\psi_{\pm}}^{\otimes N}$, respectively.

As an example of displacement estimation in a finite dimensional Hilbert space, one can consider the problem of estimation of a phase parameter $\theta$ imprinted on a quantum state $\rho_{\theta} = e^{-i\theta H}\rho_{0}e^{i\theta H}$. Let us take $H = \sum_{j=1}^{N}\sigma_{z}^{(j)}\otimes \mathbb{I}_{N-1}$ and take $\rho_{0}$ to correspond to the GHZ state $\ket{0}^{\otimes N} + \ket{1}^{\otimes N} / \sqrt{2}$. The quantum Fisher information of $\rho_{0}$ with respect to $H$ is $4N^{2}$; in fact, this is the maximum possible value of the quantum Fisher information in $(\mathbb{C}^{2})^{\otimes N}$ with respect to 1-local Hamiltonians of unit operator norm \cite{pezzesmerz}. In contrast, any product state has maximal quantum Fisher information of order $N$ over the set of such Hamiltonians. This fact suggests an ordering of superposition states based on their maximal usefulness for quantum metrology as compared to the maximal usefulness of the individual pure states which comprise the superposition. The following definition serves to characterize as ``metrologically useful'' the multimode, equal weight superposition states in separable Hilbert space that are extensively more useful for displacement estimation of a pre-defined set of self-adjoint generators than the component branches.

\vspace{0.5cm}
\textbf{Definition 1}: \textit{An equal weight quantum superposition of $q$ linearly independent pure states, $\ket{\omega} \propto  \sum_{j=1}^{q}\ket{\psi_{j}} \in (\ell^{2}(\mathbb{C}))^{\otimes N}$, is considered \textit{metrologically useful} when the following condition on the quantity $N^{rF}(\ket{\omega})$ is satisfied: }

\begin{equation}
N^{rF}(\ket{\omega}):= {\max_{H \in \mathcal{A}_{\text{1-loc.}}}\langle \omega \vert (\Delta H)^{2}\vert \omega\rangle \over {1\over q}\sum_{j=1}^{q}\max_{H \in \mathcal{A}_{\text{1-loc.}}}\langle \psi_{j} \vert (\Delta H)^{2}\vert \psi_{j} \rangle } \in \mathcal{O}(n_{\text{tot}}).
\label{eqn:nrf}
\end{equation}
\textit{where $\langle \cdot \vert (\Delta H)^{2} \vert \cdot \rangle := \langle \cdot \vert  H^{2} \vert \cdot \rangle - \langle \cdot \vert H \vert \cdot \rangle^{2}$, $n_{\text{tot}} = \langle \omega \vert \sum_{j=1}^{N}a^{\dagger}_{j}a_{j} \otimes \mathbb{I}_{N-1}\vert \omega \rangle$ is the expected total photon number,  $\mathcal{A}$ is an algebra of observables on $\ell^{2}(\mathbb{C})$, and $\mathcal{A}_{\text{1-loc.}}$ is the linear subspace of $\mathcal{A}^{\otimes N}$ in which each element is ``1-local,'' i.e., has the form $\sum_{j=1}^{N}x_{j}\otimes \mathbb{I}_{N-1}$ for $x_{j} \in \mathcal{A}$.}
\vspace{0.5cm}

The set $\mathcal{A}_{\text{1-loc.}}$ should be such that the denominator in Eq.(\ref{eqn:nrf}) is nonzero. The restriction to 1-local observables in Definition 1 allows one to use product states as a scaling standard. Specifically, given $H =\sum_{j=1}^{N}x_{j} \in \mathcal{A}_{\text{1-loc.}}$ (here, we have omitted the identity operators for clarity), and a product state $\ket{\psi}=\ket{\psi^{(1)}}\otimes \cdots \otimes \ket{\psi^{(N)}}$, it follows that:
\begin{equation}
\langle \psi \vert (\Delta H)^{2} \vert \psi \rangle \le N\text{max}_{j}\langle \psi^{(j)} \vert (\Delta x^{(j)})^{2} \vert \psi^{(j)} \rangle .
\end{equation} Hence, the variance of a measurement of a 1-local observable always scales linearly in the number of modes when the system is in a product state. A pure state $\ket{\omega}$ having the above form is metrologically useful if there exists $H$ (having the 1-local form above) such that $\langle \omega \vert (\Delta H)^{2} \vert \omega \rangle \in \mathcal{O}(Nn_{\text{tot}} \text{max}_{j,k}\langle \psi_{j}^{(k)} \vert (\Delta x^{(k)})^{2}\vert \psi_{j}^{(k)} \rangle )$.

The quantity $N^{rF}$ was originally introduced as a measure of macroscopicity for quantum superpositions in $(\mathbb{C}^{2})^{\otimes N}$ \cite{dur}; in that context, $\mathcal{A} = \mathfrak{su}(2,\mathbb{C})$ (represented by the Pauli matrices) and $n_{\text{tot.}}$ is taken to be equal to the number of modes, $N$. The notion of metrological usefulness in Definition 1 refers to the greater ultimate precision achievable in the quantum Cram\'{e}r-Rao bound when the displacement parameter is encoded in the equal weight quantum superposition state $\ket{\omega}$ compared to the ultimate precision achievable when the displacement parameter is encoded in branches $\lbrace \ket{\psi_{m}} \rbrace_{m=1}^{q}$ comprising $\ket{\omega}$. It should be noted that one can speak of a superposition state as being metrologically useful only if the algebra $\mathcal{A}$ is specified. In addition, there may be many ways to write $\ket{\omega}$ as an equal weight superposition of pure states; in this case, Definition 1 clearly refers to the metrological usefulness of $\ket{\omega}$ relative to a given decomposition of $\ket{\omega}$ into branches. In realistic parameter estimation protocols, the branch decomposition could be imposed by the preferred basis of an experiment.

In the case of a separable Hilbert space $\mathcal{H}$, the algebra $\mathcal{A}$ does not have to be represented in the von Neumann algebra $B(\mathcal{H})$. Many observables of interest, e.g., the quadrature operators and the photon number operator, are unbounded on $\mathcal{H}$ but appear in quantum optical Hamiltonians of interest to quantum metrology. However, for most quantum optical states of interest, these unbounded operators have finite second moment \cite{holevo} and so have bounded variance in these states. In particular, if an unbounded, essentially self-adjoint operator $x=x^{\dagger}$ satisfies $\langle \omega \vert x^{2} \vert \omega \rangle < \infty$ for the normalized superposition state $\ket{\omega}=\sum_{j}\ket{\psi_{j}}$, then $\langle \psi_{j} \vert x^{2} \vert \psi_{j} \rangle < \infty$ for all $j$. This feature can be used to introduce a Lie algebra for which the state $\omega$ is metrologically useful \cite{volkoff4}. $\mathcal{A}$ is then formed by taking the 1-local sums of essentially self-adjoint elements of this Lie algebra (we have here assumed a representation on $\mathcal{H}$). In order for the denominator of the expression for $N^{rF}$ to be well-defined, at least one branch $\ket{\psi_{j}}$ of $\ket{\omega}$ must not be an eigenvector of all essentially self-adjoint elements of the Lie algebra. 

A simple nontrivial example shows that $\ket{\text{ECS}_{N}^{+}(\alpha)}$ is metrologically useful when $\mathcal{A}_{\text{1-loc.}}$ is formed from observables of the Lie algebra $\mathfrak{h}_{3}=(\text{span}\lbrace a^{\dagger}, a,\mathbb{I} \rbrace , [\cdot , \cdot])$, represented as linear operators on $\ell^{2}(\mathbb{C})$ in the usual way.  Given $\alpha \in \mathbb{C}$, the even and odd coherent states $\ket{\psi_{\pm}}$ (which coincide with $\ket{\text{ECS}_{N=1}^{\pm}(\alpha)}$) exhibit a order $\vert \alpha \vert^{2}$ variance for measurements of the quadrature corresponding to the direction $\text{Arg}(\alpha)$ and exhibit squeezing in the variance of measurements of the conjugate quadrature corresponding to $\text{Arg}(\alpha)+{\pi\over 2}$ \cite{knightbuzek}. Physically, this is due to the fact that the quantized electric field is $\pi$ phase-shifted (in expectation) between $\ket{\alpha}$ and $\ket{-\alpha}$ coherent states. Explicitly, taking $\alpha \in \mathbb{R}$ and the quadratures $x^{(\theta)}$ as above, then $\langle \psi_{+} \vert (\Delta x^{(0)})^{2}\vert \psi_{+} \rangle = \alpha^{2} (1+ \tanh \alpha^{2}) + 1/2$ while $\langle \psi_{+} \vert (\Delta x^{(\pi/2)})^{2}\vert \psi_{+} \rangle =1/2 - \alpha^{2}(1-\tanh \alpha^{2})$. Thus, if $\alpha$ is purely real, the $\theta = 0$ quadrature exhibits large fluctuations, while the conjugate quadrature fluctuates just below the vacuum level. Since the only observables arising from $\mathfrak{h}_{3}$ are the oscillator quadratures and the identity, it is clear that an observable $\sum_{j=1}^{N}z_{j}a^{\dagger}_{j} + \overline{z}_{j}a_{j}$ exists that exhibits variance on the order of $N^{2}\alpha^{2}\max_{j}\vert z_{j}\vert^{2}$ in $\ket{\text{ECS}_{N}^{+}(\alpha)}$. On the other hand, since every quadrature $x^{(\theta)}$ exhibits variance of 1/2 in the coherent state $\ket{\pm \alpha}$, any 1-local observable $\sum_{j=1}^{N}z_{j}a^{\dagger}_{j} + \overline{z}_{j}a_{j}$ has variance of order $N\max_{j}\vert z_{j} \vert^{2}$ in $\ket{\pm \alpha}^{\otimes N}$. Using Definition 1, we see that taking $\mathcal{A}_{\text{1-loc.}}$ to be composed of observables from $\mathfrak{h}_{3}$ allows $\ket{\text{ECS}_{N}^{\pm}(\alpha)}$ to be considered metrologically useful. In particular, by taking $z_{j} = x \in \mathbb{R}$, $\ket{\text{ECS}_{N}^{\pm}(\alpha)}$ are metrologically useful for estimation of global amplitude displacements $\bigotimes_{j=1}^{N} D_{j}(x)$ of $N$-mode oscillators. The estimation of arbitrary local displacements in the complex plane comprises a multiparameter ($2N$ real parameters) estimation task \cite{parismultiparam}. It is an interesting problem whether a measure analogous to $N^{rF}$ can be used to identify Schr\"{o}dinger cat states as a resource for parameter estimation of more general quantum dynamics.

It should be noted that $\ket{\text{ECS}_{N}^{\pm}(\alpha)}$ is not metrologically useful when $\mathcal{A}_{\text{1-loc.}}$ is composed of observables from the oscillator Lie algebra $\mathfrak{h_{4}} = (\text{span}\lbrace a^{\dagger}a, a^{\dagger}, a,\mathbb{I} \rbrace , [\cdot , \cdot])$ instead of $\mathfrak{h}_{3}$. This is because the 1-local photon number operator $\sum_{j=1}^{N}a^{\dagger}_{j}a_{j}$ exhibits extensive variance $N\vert \alpha \vert^{2}$ in the coherent states $\ket{\pm \alpha}^{\otimes N}$, so that the ratio in Definition 1 exhibits linearly scaling with $N$, the number of modes, and not the total number of photons.

We now detail the argument that $\ket{\text{HCS}_{N}^{\pm}(\alpha)}$ are metrologically useful when the algebra $\mathcal{A}$ is the Lie algebra of observables of $\mathfrak{sl}(2,\mathbb{C})$. Algebraically, it is simpler to show this fact for a closely related hierarchical cat state. By returning to (\ref{eqn:hcs}) and taking $\ket{\phi }=\ket{\psi_{+}}$ and $U = e^{i\pi a^{\dagger}a/2}(\ket{\psi_{+}}\bra{\psi_{-}}+\ket{\psi_{-}}\bra{\psi_{+}})$, the following state is produced:
\begin{equation}
\ket{\Omega(\alpha)} = {1\over \sqrt{2}}\left( \left( {\ket{\alpha}+\ket{-\alpha}\over \sqrt{2+2e^{-2\vert \alpha \vert^{2}}}}\right)^{\otimes N} + \left( {\ket{i\alpha}-\ket{-i\alpha}\over \sqrt{2-2e^{-2\vert \alpha \vert^{2}}}} \right)^{\otimes N} \right).
\end{equation}

Consider the 1-local Hamiltonian $\sum_{j=1}^{N}(\overline{z}a^{(j)2} + za^{\dagger (j)2} ) \otimes \mathbb{I}_{N-1}$ as would describe two-photon parametric downconversion into $N$ modes, each with classical pumping amplitude $z\in \mathbb{C}$. Because $a^{2}\ket{\psi_{+}} = \alpha^{2}\ket{\psi_{+}}$ and $a^{2}e^{i\pi a^{\dagger}a/2}\ket{\psi_{-}} = -\alpha^{2}e^{i\pi a^{\dagger}a/2}\ket{\psi_{-}}$, it is clear that for any states $\ket{\xi_{1}}$, $\ket{\xi_{2}}\in \text{span}_{C}\lbrace \ket{\psi_{+}},e^{i\pi a^{\dagger}a/2}\ket{\psi_{-}}\rbrace$, the following Pauli matrix/two-photon quadrature duality holds:
\begin{equation}
\langle \xi_{1} \vert \sigma_{z}  \vert \xi_{2} \rangle = \langle \xi_{1}\vert {1\over \alpha^{2}}(a^{2}+a^{\dagger 2}) \vert \xi_{2}\rangle
\end{equation}
where $\sigma_{z} = \ket{\psi_{+}}\bra{\psi_{+}} - e^{i\pi a^{\dagger}a/2}\ket{\psi_{-}}\bra{\psi_{-}}e^{-i\pi a^{\dagger}a/2}$ is the appropriate Pauli matrix in $\text{span}_{C}\lbrace \ket{\psi_{+}},e^{i\pi a^{\dagger}a/2}\ket{\psi_{-}}\rbrace$. From this, it is clear that the variance of $\sum_{j=1}^{N}(\overline{z}a^{(j)2} + za^{\dagger (j)2} ) \otimes \mathbb{I}_{N-1}$ in $\ket{\Omega(\alpha)}$ should be of order $N^{2}\vert z \vert^{2}\vert\alpha\vert^{4}$. In fact, for $\alpha \in \mathbb{R}$ the variance is:
\begin{eqnarray}
&{}&4N^{2}\text{Re}(\overline{z}\alpha^{2})^{2} + {N\over 2}( 4\text{Re}(\overline{z}^{2}\alpha^{4}) -8\text{Re}(\overline{z}\alpha^{2})^{2} \nonumber \\&+& 4\vert z\vert^{2}\vert\alpha\vert^{2}(\tanh \alpha^{2} + \coth \alpha^{2}) + 2\vert z\vert^{2} )
\end{eqnarray}
which is on the order of $N^{2}\vert z \vert^{2}\vert\alpha\vert^{4}$ for $\text{Arg}z=2\text{Arg}\alpha$. In addition, the variance of $\sum_{j=1}^{N}(\overline{z}a^{(j)2} + za^{\dagger (j)2} ) \otimes \mathbb{I}_{N-1}$ in the product states $\ket{\psi_{+}}^{\otimes N}$ or $(e^{i\pi a^{\dagger}a/2}\ket{\psi_{-}})^{\otimes N}$ is at most of order $N\vert z \vert^{2}\vert \alpha \vert^{2}$, as can easily be verified. The final step in finding a minimal algebra $\mathcal{A}$ which allows $\ket{\Omega(\alpha)}$ to be metrologically useful is to append the element $(1/2)a^{\dagger}a + 1/4$ to the set $\lbrace a^{\dagger 2}/2,a^{2}/2 \rbrace$ and check that the 1-local observable given by, e.g., $\sum_{j=1}^{N}a^{\dagger}_{i}a_{i}$ does not exhibit fluctuations in either of the branch states $\ket{\psi_{+}}^{\otimes N}$ or $(e^{i\pi a^{\dagger}a/2}\ket{\psi_{-}})^{\otimes N}$ scaling as $\vert \alpha \vert^{4}$. If the 1-local photon number operator were to exhibit such fluctuations, then the ratio on the left hand side of Eq.(\ref{eqn:nrf}) would lose the property of scaling with the total expected number of photons in $\ket{\Omega(\alpha)}$. It is simple to verify that the 1-local photon number operator exhibits variance of order $N\vert \alpha \vert^{2}$ in these product states and hence, $\ket{\Omega(\alpha)}$ is metrologically useful when $\mathcal{A} = \mathfrak{sl}(2,\mathbb{C}) := (\text{span}_{\mathbb{C}}\lbrace (1/2)a^{\dagger}a + 1/4, a^{\dagger 2}/2,a^{2}/2 \rbrace , [\cdot , \cdot])$.

In particular, when $z \in \mathbb{R}$ the calculation above shows that $\ket{\text{HCS}_{N}^{\pm}(\alpha)}$ are metrologically useful for displacement estimation, where the displacement parameter now corresponds to the global squeezing amplitude $z \in \mathbb{R}$. It is intriguing that while some superpositions of $\mathfrak{sp}(N,\mathbb{C})$ Barut-Girardello coherent states do exhibit squeezing, the $\ket{\text{HCS}_{N}^{+}(\alpha)}$ state does not; in addition, the product states comprising each of the branches of $\ket{\text{HCS}_{N}^{+}(\alpha)}$ exhibit negligible squeezing if $\vert \alpha \vert^{2} > 1$.  

However, if squeezed states and their superpositions are available, one may wonder if there exist other types of hierarchical cat states having $N^{rF}$ scaling exponentially in a squeezing parameter when the observables of $\mathcal{A}_{\text{1-loc.}}$ are taken from $\mathfrak{sl}(2,\mathbb{C})$. Indeed, it is known that squeezed states provide a higher precision in the estimation of a single mode squeezing parameter than coherent states \cite{chiribella}.  The following hierarchical cat state, having branches composed of superpositions of ideal squeezed states, allows for such scaling:
\begin{equation}
{1\over \sqrt{2}}\left( \left( {(D(\alpha)+D(-\alpha))S(w) \over \sqrt{2+2\text{exp}(-2\alpha^{2}e^{2w})}} \ket{0}\right)^{\otimes N} +   \left( {(D(i\alpha)-D(-i\alpha))S(w) \over \sqrt{2-2\text{exp}(-2\alpha^{2}e^{2w})}} \ket{0}\right)^{\otimes N}\right)
\end{equation}
where we have assumed $\alpha , w \in \mathbb{R}_{>0}$ and taken $S(w) := e^{(1/2)(\overline{w}a^{2}-wa^{\dagger 2})}$ as the unitary squeezing operator. For such $\alpha$ and $w$, the identity $D(\alpha)S(w)=S(w)D(\alpha e^{w})$ holds and so the above hierarchical cat state can be rewritten as $S(w)^{\otimes N}\ket{\Omega(\alpha e^{w})}$. The $N^{rF}$ value of $S(w)^{\otimes N}\ket{\Omega(\alpha e^{w})}$ exhibits the same scaling as the $N^{rF}$ value for $\ket{\Omega(\alpha e^{w})}$, i.e., of order $N \alpha^{2}e^{2w}$, because $S(w)$ acts on $\mathfrak{sl}(2,\mathbb{C})$ via the adjoint action. As a final remark, we point out that the coherent states $\pm \ket{\alpha}$ are minimum uncertainty states for the Heisenberg uncertainty relation for observables of $\mathfrak{h}_{3}$ while the even/odd coherent states $\ket{\psi_{\pm}}$ are minimum uncertainty states for the generalized uncertainty relation for observables of $\mathfrak{su}(2,\mathbb{C})$ \cite{brif}. This is not a surprising coincidence, as the definition of metrological usefulness (Definition 1) requires that the maximal uncertainty in the product states comprising the branches of a state having the form (\ref{eqn:ecsgen}) be extensively smaller than the maximal uncertainty in the multimode superposition state.

Thus far, the discussion of Definition 1 has been mainly mathematical. It is useful to mention that from a basic physical perspective, the problem of determining the precision of optimal estimation of a global real displacement parameter is equivalent to determing the energy-time uncertainty in a given quantum state. Therefore Definition 1 can be reinterpreted from a physical perspective by stating that $\ket{\omega}$ is metrologically useful with respect to Hamiltonians $H\in \mathcal{A}_{\text{1-loc.}}$ if its maximal decay rate (i.e., minimal time $t$ for which $e^{-iHt}\ket{\omega}$ becomes distinguishable from $\ket{\omega}$) is extensively greater than the maximal decay rates of the branch states $\lbrace\ket{\psi_{j}}\rbrace_{j=1}^{q}$, i.e., if $\ket{\omega}$ is extensively more sensitive to evolution generated by $H$ as compared to the branches. In the particular case of 
$\ket{\text{HCS}_{N}^{+}(\alpha)}$ ($\ket{\text{ECS}_{N}^{+}(\alpha)}$), one can say qualitatively that its metrological usefulness arises simply because its intermode quantum coherence causes the squeezing operation (displacement operation) to change it more drastically than the product states $\ket{\psi_{\pm}}^{\otimes N}$ ($\ket{\pm\alpha}^{\otimes N}$) considered independently.

\subsection{Entanglement entropy}
Because of the orthogonality of the branches, the mode entanglement structure of $\ket{\text{HCS}_{2}^{\pm}}$ is the same as that of the GHZ states in $(\mathbb{C}^{2})^{\otimes 2}$. Hence, $\ket{\text{HCS}_{2}^{\pm}}$ exhibits maximal entanglement entropy in the $(\mathbb{C}^{2})^{\otimes 2}$ subspace spanned by $\lbrace \ket{e_{i}}\otimes \ket{e_{j}} \rbrace_{i,j = 1,2}$ introduced in Section \ref{sec:intro}. In particular, $\ket{\text{HCS}_{2}^{\pm}(\alpha)}$ exhibits greater mode entanglement than the subset of entangled coherent states that cannot be expressed in the form (\ref{eqn:hcs}). However, the hierarchical photonic superpositions are not maximally entangled states of $\ell^{2}(\mathbb{C}) \otimes \ell^{2}(\mathbb{C})$. The dissipative dynamics of the entanglement entropy of entangled coherent states was studied in Refs.\cite{jeongbell,zagury,jeongkim}.

\begin{figure}[t!]
\includegraphics[scale=0.5]{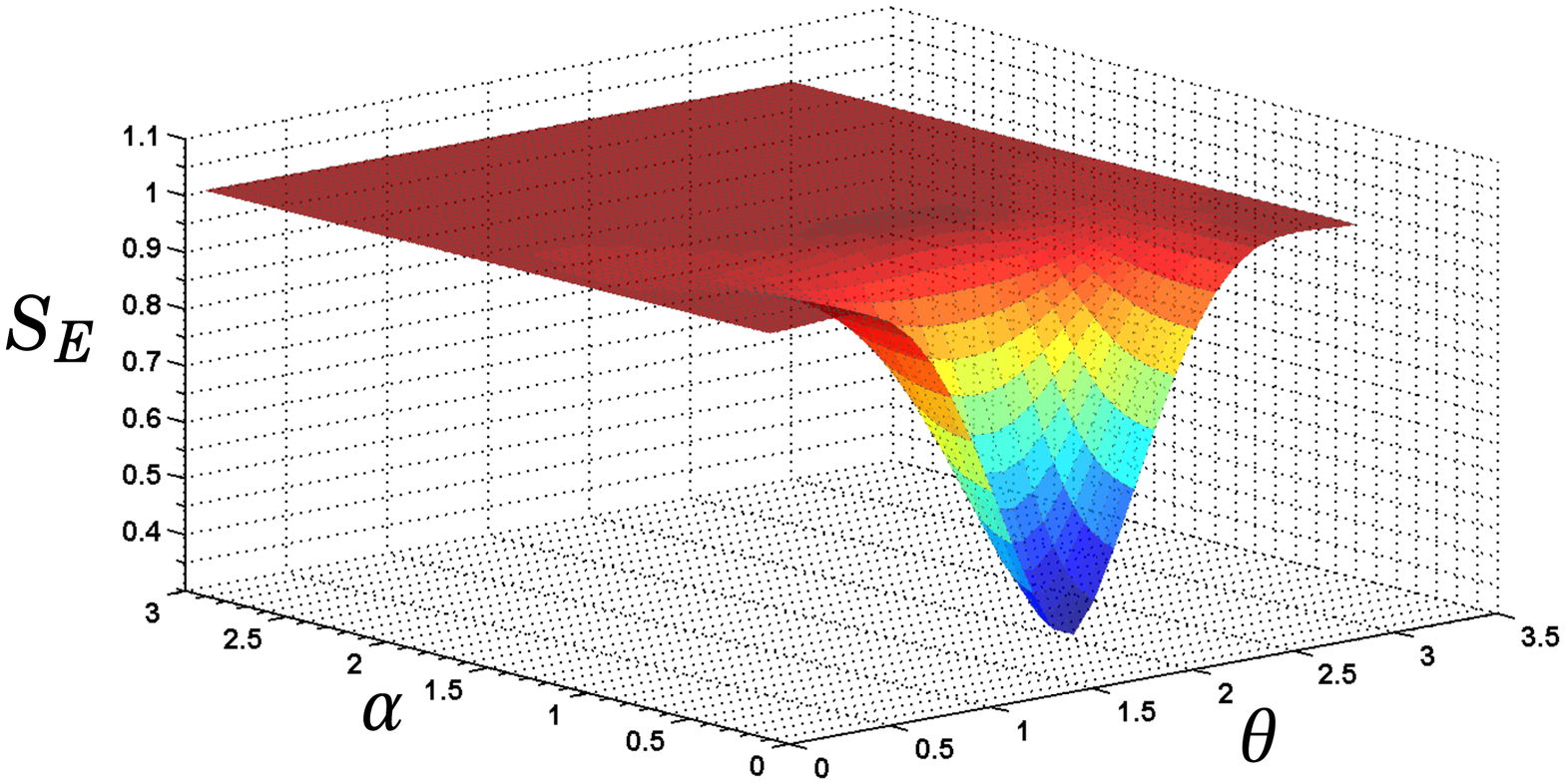}
\caption{The entanglement entropy of $U(\theta)\ket{\text{HCS}_{N=2}^{+}(\alpha)}$, $\alpha \in [0.2,3.0]$ and $\theta \in [0.1,\pi -0.1]$.\label{fig:entangleentbeam}}
\end{figure}

It is known that a nonclassical product state incident on a beam splitter does not necessarily generate entanglement between the output modes \cite{kimbuzek}. In fact, a 50:50 beam splitter destroys the entanglement of a two-mode squeezed state \cite{agarwalbook}. It is easy to see that a beam splitter described by the unitary operation $B(\theta)=e^{i\theta /2(a_{1}^{\dagger}a_{2}+a_{2}^{\dagger}a_{1})}$ acting on two input photonic modes maps $\ket{\text{HCS}_{2}^{+}(\alpha)}$ to the state
\begin{eqnarray}
&{}& {1\over \sqrt{2}}({1\over 1-e^{-4\vert \alpha \vert^{2}}}(\ket{\alpha e^{i\theta/2}}\otimes \ket{\alpha e^{i\theta/2}} + \ket{-\alpha e^{i\theta/2}}\otimes \ket{-\alpha e^{i\theta/2}}) \nonumber \\ &-& {1\over 2 \sinh(2\vert \alpha \vert^{2})}(\ket{\alpha e^{-i\theta/2}}\otimes \ket{-\alpha e^{-i\theta/2}} + \ket{-\alpha e^{-i\theta/2}}\otimes \ket{\alpha e^{-i\theta/2}} ) ).
\end{eqnarray}
For moderately large $\vert \alpha \vert$, the exponentially decaying term becomes negligible and one is left with an entangled coherent state in the output modes of the beam splitter. Hence, the beam splitter does not destroy the entanglement of $\ket{\text{HCS}_{2}^{+}(\alpha)}$ for any values of the transmission and reflection amplitudes. The exact entanglement entropy of $B(\theta) \ket{\text{HCS}_{2}^{+}(\alpha)}$, calculcated as the von Neumann entropy of the reduced density matrix, is shown in Fig.\ref{fig:entangleentbeam} for a range of real $\alpha$ and $\theta$. Except for low-power  ($\alpha \lesssim 1$) $\ket{\text{HCS}_{2}^{+}(\alpha)}$ states, maximum entanglement entropy is maintained throughout the range of transmission amplitudes of the beam splitter.

\begin{figure}[t!]
\includegraphics[scale=0.5]{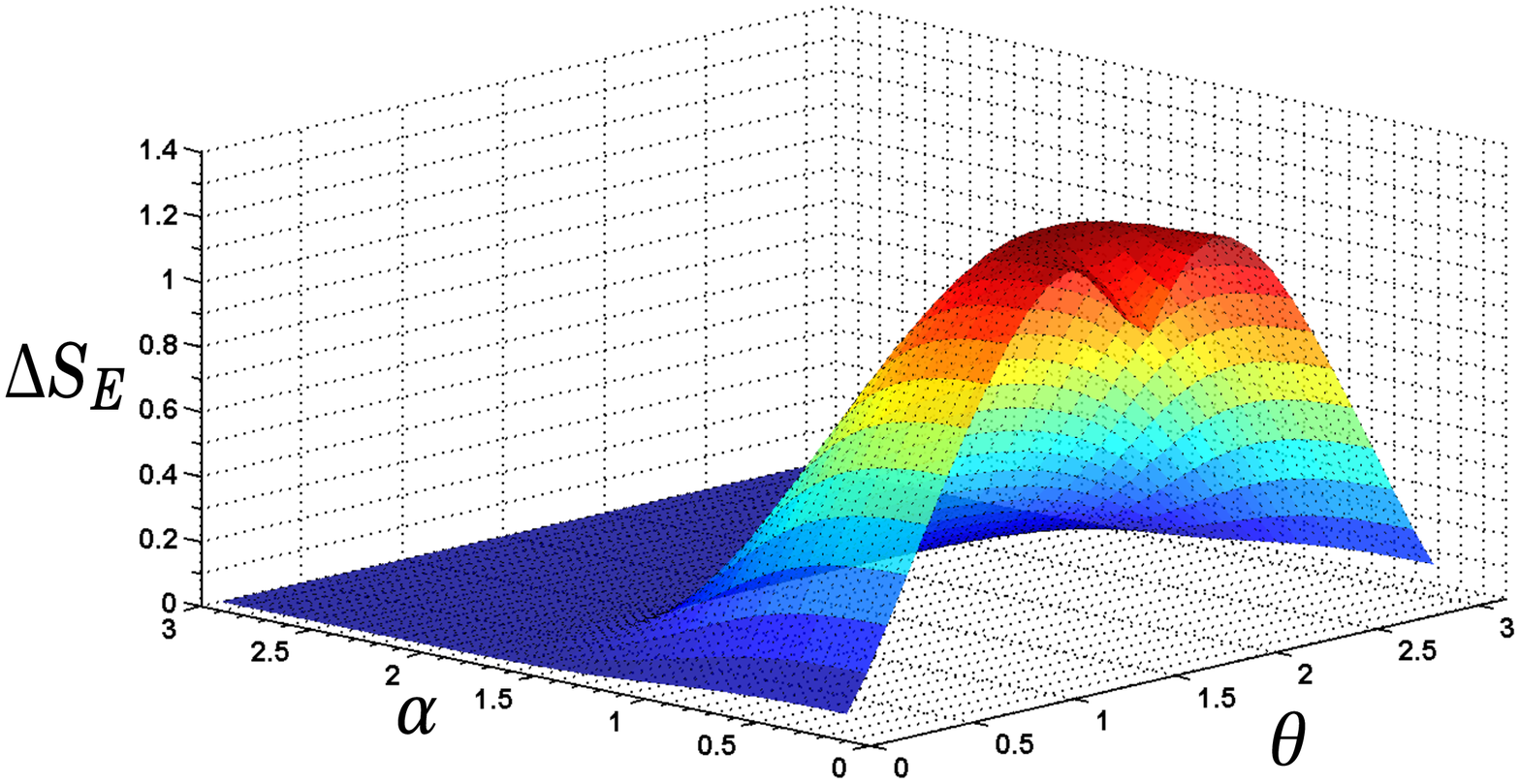}
\caption{The entanglement entropy fluctuations of $U(\theta)\ket{\text{HCS}_{N=2}^{+}(\alpha)}$, $\alpha \in [0.2,3.0]$ and $\theta \in [0.1,\pi -0.1]$.\label{fig:fluct}}
\end{figure}

The quantification of entanglement in terms of an entropic quantity naturally leads to questions about its fluctuations. Entanglement fluctuations can be interpreted as the root variance of a measurement of the entanglement Hamiltonian \cite{fluctentang}, i.e., in terms of the reduced density matrix $\rho_{A}$ of a pure state of $\mathcal{H}_{A}\otimes \mathcal{H}_{B}$ it is given by the expression $\Delta S_{E} := \sqrt{\text{tr}(\rho_{A}H_{E}^{2}) - \text{tr}(\rho_{A}H_{E})^{2}}$ where $H_{E}:=-\log_{2}\rho_{A}$. We show the entanglement fluctuation of $B(\theta) \ket{\text{HCS}_{2}^{+}(\alpha)}$ in Fig.\ref{fig:fluct}.

\begin{figure}[t!]
\includegraphics[scale=0.5]{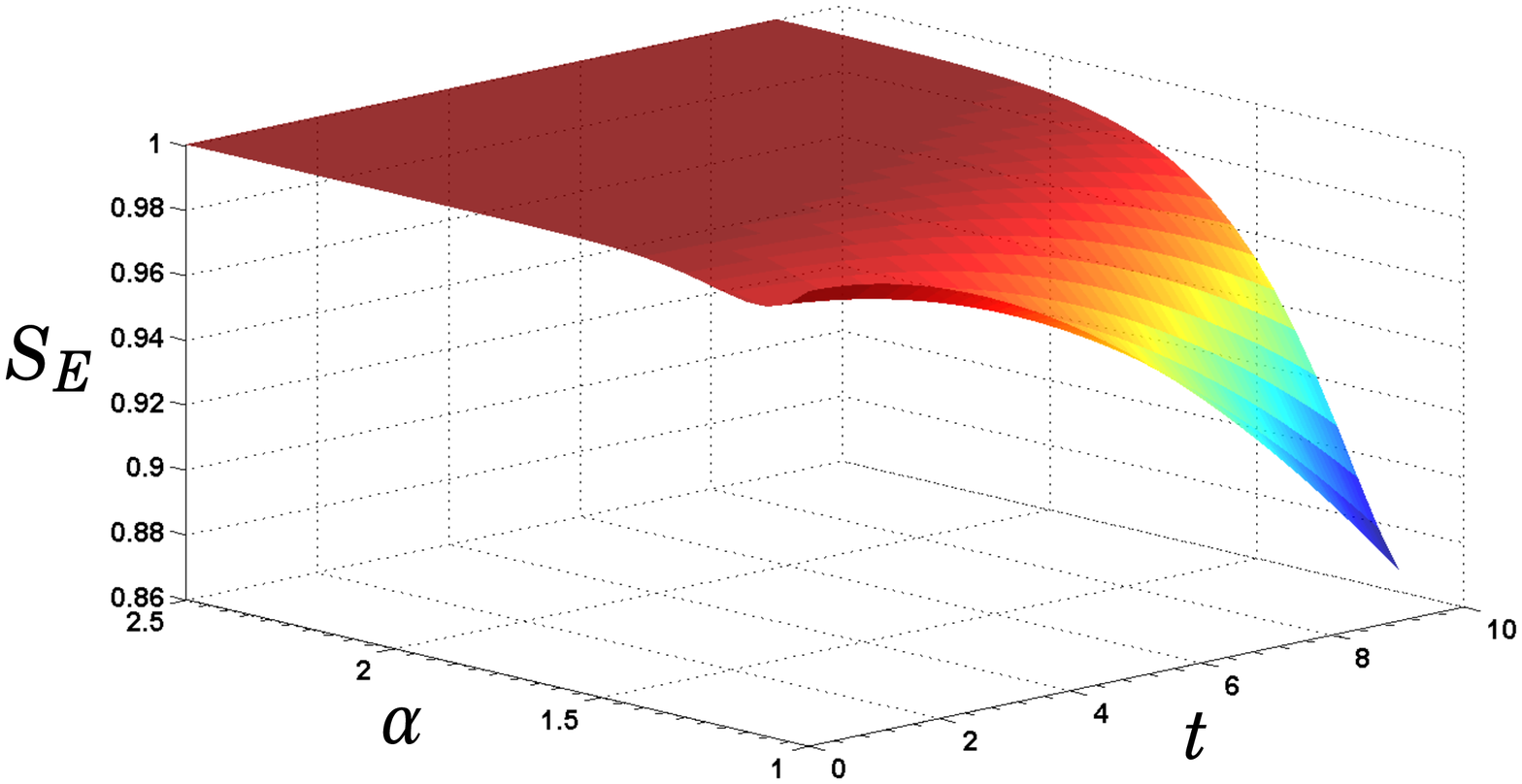}
\caption{The entanglement entropy of the image of $\ket{\text{HCS}_{N=2}^{+}(\alpha)}$ under independent amplitude damping for $\Gamma = 0.1$, $\alpha \in [1,2.5]$ and $t \in [0,9]$.\label{fig:nonhamentang}}
\end{figure}

We now show that in the Bell basis $\lbrace \ket{\text{HCS}_{2}^{+}(\alpha)} , \ket{\text{HCS}_{2}^{-}(\alpha)} , \ket{\text{ECS}_{2}^{-}(\alpha)} , (e^{i\pi a^{\dagger}a}\otimes \mathbb{I})\ket{\text{ECS}_{2}^{-}(\alpha)} \rbrace $, the hierarchical cat states comprise, in some sense, the most stable entanglement resource. We consider each mode coupled independently to a zero temperature bath of photons, each bath having absorption rate $\Gamma$, with the non-Hamiltonian part of the evolution given by
\begin{equation}
\rho ' = {\Gamma \over 2}\sum_{j=1}^{2}[a,\rho(t)a^{\dagger}] + [a\rho , a^{\dagger}] .\label{eqn:lindblad}
\end{equation}
This is the case of (independent) Lindbladian amplitude damping. For an initial state $\rho(t=0)$ an entangled coherent state, it follows from the well-known solution of the amplitude damping master equation \cite{barnett} that the $t\rightarrow \infty$ asymptotic is unentangled. In contrast, $\ket{\text{HCS}_{2}^{\pm}(\alpha)}$ maintain nearly maximal entanglement entropy $S_{E}$ throughout the non-Hamiltonian evolution as long as $\vert \alpha \vert^{2} \gtrsim 1$, as seen in Fig. \ref{fig:nonhamentang}. 

The entanglement entropy for this state was calculated from the analytical expression, which we omit. It should be noted that for $\vert \alpha \vert \lesssim 1$, the entanglement entropy decays with time, but is still substantial for $t \gtrsim \Gamma^{-1}$. The persistence of entanglement under the amplitude damping map can be simply seen by considering a limiting model. By taking the low-power $\alpha \rightarrow 0 $ limit, it is clear that $\ket{\text{HCS}_{2}^{+}(\alpha)}$ exhibits an inner product of magnitude 1 with the state
\begin{equation}
{1\over \sqrt{2}}(\ket{0}^{\otimes 2} + \ket{1}^{\otimes 2}),
\label{eqn:focksuper}
\end{equation}
i.e., a superposition of two-mode vacuum and the product Fock state $\ket{1}\otimes \ket{1}$. Under the amplitude damping map defined in Eq.(\ref{eqn:lindblad}), the state (\ref{eqn:focksuper}) evolves to
\begin{eqnarray}
&&{1\over 2}(\ket{0}\bra{0}\otimes \ket{0}\bra{0} + e^{-2\Gamma t}(\ket{0}\bra{1}\otimes \ket{0}\bra{1} + \ket{1}\bra{0}\otimes \ket{1}\bra{0}) \nonumber \\
&+& (e^{-2\Gamma t}\ket{1}\bra{1} + (1-e^{-2\Gamma t})\ket{0}\bra{0})\otimes (e^{-2\Gamma t}\ket{1}\bra{1} + (1-e^{-2\Gamma t})\ket{0}\bra{0}) ).
\end{eqnarray}
Taking the partial trace to form $\rho_{1}(t)$, one computes $\lim_{t\rightarrow \infty} -\text{tr}(\rho_{1}(t)\log_{2}\rho_{1}(t)) = 1$. The robustness of the entanglement entropy under amplitude damping exhibited for large $\vert \alpha \vert$ is proven by considering the $\vert \alpha \vert \rightarrow \infty$ asymptotics. An explicit calculation shows that for any finite $\vert \alpha \vert$, $\lim_{t\rightarrow \infty}S_{E} = 0$, whereas $\lim_{t\rightarrow \infty}\lim_{\vert \alpha \vert \rightarrow \infty}S_{E} = 1$. The increased stability of the entanglement of hierarchical cat states to local amplitude damping (relative to the entanglement of entangled coherent states) makes these states desirable targets for quantum optical state engineering and optical quantum communication.

\section{Generation of $\ket{\text{HCS}_{2}^{+}(\alpha)}$}

In this section, we limit ourselves to proposals for experimental generation of the two-mode hierarchical superposition state $\ket{\text{HCS}_{2}^{+}(\alpha)}$ because the main difficulties are already present in this case. For all of the proposals we describe, a generalization to $N>2$ requires the experimenter to overcome a linear (with $N$) increase in errors associated with imperfect implementation of the required unitary operations, in addition to the usual problem of decoherence due to photon losses.

\begin{figure}[t!]
\includegraphics[scale=0.7]{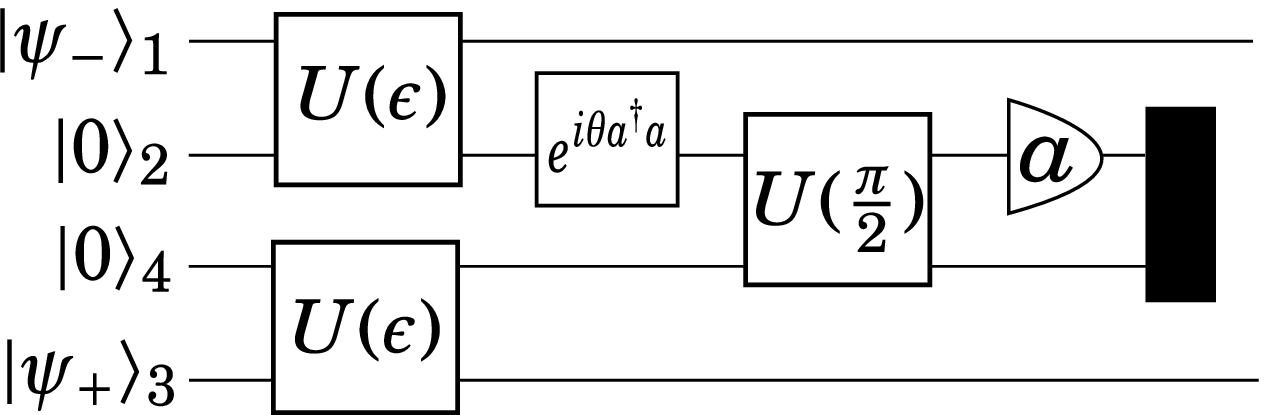}
\caption{Quantum circuit diagram for the transformation $\ket{\psi_{-}}_{1}\otimes \ket{0}_{2}\otimes \ket{\psi_{+}}_{3}\otimes \ket{0}_{4} \rightarrow {1\over \sqrt{2}}\left( \ket{\psi_{+}}_{1}\otimes \ket{\psi_{+}}_{3} + e^{-i\theta} \ket{\psi_{-}}_{1}\otimes \ket{\psi_{-}}_{3} \right)$.\label{fig:coherentphotonlossfig}}
\end{figure}

If an experimenter has access to arbitrary unitary operations in the $\mathbb{C}^{2}$ sub-Hilbert space spanned by $\ket{\pm \alpha}$ over a range of amplitudes $\alpha$, then $\ket{\text{HCS}_{2}^{+}(\alpha)}$ can be readily generated. Specifically, one applies the ``$\pi/2$'' (or ``50:50'') beam splitter $U_{12}(\pi/2)=e^{({1\over 2}){\pi\over 2}(a_{1}^{\dagger}a_{2}-a_{2}^{\dagger}a_{1})}$ to the product state $\propto (\ket{\sqrt{2}\alpha}_{1} -\ket{-\sqrt{2}\alpha}_{1})\otimes \ket{0}_{2}$ to produce the state $\propto \ket{\alpha}_{1}\ket{-\alpha}_{2} - \ket{-\alpha}_{1}\ket{\alpha}_{2}$. Applying the phase shift $e^{i\pi a^{\dagger}_{2}a_{2}}$ produces the Bell state $(1/\sqrt{2})(\ket{\psi_{+}}\ket{\psi_{-}} +\ket{\psi_{-}}\ket{\psi_{+}})$. Applying $\sigma_{x} = \ket{\psi_{+}}\bra{\psi_{-}} + \ket{\psi_{-}}\bra{\psi_{+}}$ on mode 2 produces $\ket{\text{HCS}_{2}^{+}(\alpha)}$; alternatively, $\ket{\text{HCS}_{2}^{-}(\alpha)}$ is produced (up to a global phase) conditional on the application of annihilation operator $\mathbb{I}\otimes a_{2}$ to the above Bell state. Along these lines, the method of Ref.\cite{ourj} for preparing entangled coherent states by a coherent photon loss may be modified in a simple way to produce the following family of states:
\begin{equation}
{1\over \sqrt{2}}\left( \ket{\psi_{+}}^{\otimes 2} \pm e^{-i\theta}\ket{\psi_{-}}^{\otimes 2}\right).
\label{eqn:hcsphase}
\end{equation}
The method is based on the observation that a coherent photon loss can generate photonic HCS from a product of single-mode Schr\"{o}dinger cat states. For example, the above HCS state is equivalent (in projective Hilbert space) to $(a-be^{-i\theta})(\ket{\psi_{-}}_{a}\otimes \ket{\psi_{+}}_{b})$. The implementation of the coherent photon loss via a linear quantum optical circuit is shown in Fig.\ref{fig:coherentphotonlossfig}. 
In detail, we append vacuum modes to the tensor product $\ket{\psi_{-}}\otimes \ket{\psi_{+}}$ to form the initial state  $\ket{\psi_{-}}_{1}\otimes \ket{0}_{2} \otimes \ket{\psi_{+}}_{3}\otimes \ket{0}_{4}$. $U_{ij}(\epsilon) := e^{{1\over 2}\epsilon (a^{\dagger}_{i}a_{j} - a^{\dagger}_{j}a_{i})}$ is a 50:50 beam splitter with $\epsilon \ll 1$, i.e., the beam splitter is highly transmissive for mode $i$. Applying $U_{24}(\pi/2)e^{i\phi a^{\dagger}_{2}a_{2}}U_{12}(\epsilon)U_{34}(\epsilon)$ to the initial state produces:
\begin{eqnarray}
&{}&\ket{\alpha\cos\epsilon}_{1}\otimes \ket{-\alpha{e^{i\phi}+1 \over \sqrt{2}}\sin\epsilon}_{2}\otimes \ket{\alpha{e^{i\phi}-1 \over \sqrt{2}}\sin\epsilon}_{4} \otimes \ket{\alpha\cos\epsilon}_{3} \nonumber \\ &{+}&
\ket{\alpha\cos\epsilon}_{1}\otimes \ket{-\alpha{e^{i\phi}-1 \over \sqrt{2}}\sin\epsilon}_{2}\otimes \ket{\alpha{e^{i\phi}+1 \over \sqrt{2}}\sin\epsilon}_{4} \otimes \ket{-\alpha\cos\epsilon}_{3} \nonumber \\ &{-}&
\ket{-\alpha\cos\epsilon}_{1}\otimes \ket{\alpha{e^{i\phi}-1 \over \sqrt{2}}\sin\epsilon}_{2}\otimes \ket{-\alpha{e^{i\phi}+1 \over \sqrt{2}}\sin\epsilon}_{4} \otimes \ket{\alpha\cos\epsilon}_{3} \nonumber \\ &{-}&
\ket{-\alpha\cos\epsilon}_{1}\otimes \ket{\alpha{e^{i\phi}+1 \over \sqrt{2}}\sin\epsilon}_{2}\otimes \ket{-\alpha{e^{i\phi}-1 \over \sqrt{2}}\sin\epsilon}_{4} \otimes \ket{-\alpha\cos\epsilon}_{3} .
\end{eqnarray}
The coherent photon loss is now implemented by photodetection on mode 2, modeled by application of the annihilation operator $a_{2}$. In the final step, one traces over the modes 2 and 4. In $\epsilon \rightarrow 0$ limit, the ``+'' state of expression (\ref{eqn:hcsphase}) is produced; if photodetection is carried out on mode 4 instead of mode 2, then the ``-'' state is produced. For initial superpositions $\ket{\psi_{\pm}}$ having large $\vert \alpha \vert^{2}$, $\epsilon$ must be concomitantly decreased to maintain high fidelity of the output state to $\ket{\text{HCS}_{2}^{\pm}(\alpha)}$. The decrease in $\epsilon$ necessarily increases noise in the photodetection process. In addition, for large $\vert \alpha \vert^{2}$, it is vital to generate the initial product state $\ket{\psi_{-}}_{1}\otimes \ket{\psi_{+}}_{3}$ with high fidelity. The next method that we discuss readily satisfies this requirement.

The experimental generation of single-mode photonic Schr\"{o}dinger cat states $\ket{\psi_{\pm}}$ via dispersive interaction between the monochromatic electromagnetic field and a superconducting two-level system \cite{schoelkopfcat} or Rydberg atom \cite{harochedavidovich} provides some clues toward feasible methods for preparation of photonic HCS. To extend these protocols to the many-mode case, one must effectively entangle the field states of spatially separated resonating cavities. For example, it has been proposed to generate entangled coherent states by sequential coupling of a Rydberg atom to two microwave cavities \cite{harocheecs}. In general, proposals for creating entangled field states involve coupling the field modes to easily controllable, low-dimensional quantum systems.

A simple scheme for generating $\ket{\text{HCS}_{2}^{+}(\alpha)}$ from a tensor product of even coherent states $\ket{\psi_{+}}\otimes \ket{\psi_{+}}$ is as follows: \begin{equation}
\ket{\psi_{+}}\otimes \ket{\psi_{+}} \xrightarrow{H\otimes \mathbb{I}}  {1\over \sqrt{2}}( \ket{\psi_{+}} + \ket{\psi_{-}})\otimes \ket{\psi_{+}} \xrightarrow{\text{CNOT}} \ket{\text{HCS}_{2}^{+}(\alpha)}
\label{eqn:formation}\end{equation}
where $H := 1/\sqrt{2}(\sigma_{x} + \sigma_{z})$ is the Hadamard gate in the subspace $\mathcal{K}$ spanned by orthonormal basis of even/odd coherent states and $\text{CNOT}:= \ket{\psi_{+}}\bra{\psi_{+}}\otimes \mathbb{I} + \ket{\psi_{-}}\bra{\psi_{-}}\otimes \sigma_{x}$ is the conditional $\sigma_{x}$ operation on the second field mode. To implement the Hadamard operation, it is sufficient to generate the following superposition: \begin{eqnarray} H\ket{\psi_{+}} &=& {1\over \sqrt{2}}\left( \ket{\psi_{+}} + \ket{\psi_{-}} \right) \nonumber\\ &=& {\sqrt{1+e^{-2\alpha^{2}}} + \sqrt{1-e^{-2\alpha^{2}}}  \over 2\sqrt{1-e^{-4\alpha^{2}}}}\ket{\alpha} + {\sqrt{1+e^{-2\alpha^{2}}} - \sqrt{1-e^{-2\alpha^{2}}}  \over 2\sqrt{1-e^{-4\alpha^{2}}}}\ket{-\alpha}.\end{eqnarray}
Arbitrary superpositions of photonic coherent states $\ket{\pm \alpha}$ can be generated by a dispersive coupling between a coherent microwave field $\ket{\alpha}$ and a transmon qubit if the transmon qubit can be prepared in an arbitrary pure state of $\mathbb{C}^{2}$ \cite{schoelkopfcat}. In addition, it has been proposed to generate parametrically tuning \cite{albertkrastonov}. It is worth noting that $H\ket{\psi_{+}}$ is an eigenvector of the operator $\ket{\alpha}\bra{\alpha} - \ket{-\alpha}\bra{-\alpha}$, which is proportional to the observable corresponding to the measurement which optimally detects $\ket{\alpha}$ or $\ket{-\alpha}$ (in the sense of quantum binary distinguishability problem with equal $\textit{a priori}$ probabilities and Bayes' cost criterion \cite{holevoqbook})  with maximal probability of success. The pure states $H\ket{\psi_{\pm}}$ have been studied for their role in optimal detection of coherent states $\ket{\pm \alpha}$, i.e., the ``binary phase shift key'' \cite{holevohirota}. 

The CNOT gate in the scheme (\ref{eqn:formation}) is more difficult to engineer than the Hadamard gate because it requires not only a large intramode coherence time for the even and odd coherent states, but also a large intermode coherence between two microwave cavities. However, if two transmon qubits can be prepared in a maximally entangled (i.e., GHZ) state of $(\mathbb{C}^{2})^{\otimes 2}$ and independently coupled to spatially separated photonic modes of microwave cavities via a dispersive interaction, this CNOT gate can be implemented. We now provide the details for factoring the unitary operator corresponding to the CNOT gate into easily implementable unitary operations on the field/qubit and qubit/qubit subsystems.

\begin{figure}[t!]
\includegraphics[scale=0.7]{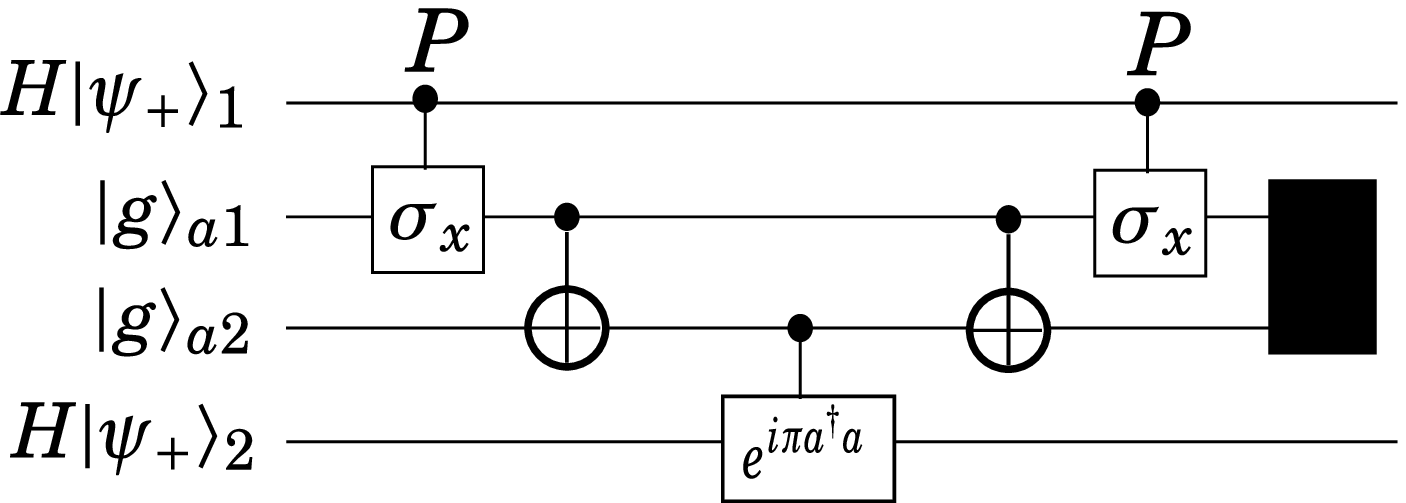}
\caption{Quantum circuit diagram for the transformation $H\ket{\psi_{+}}_{1}\otimes H\ket{\psi_{+}}_{2} \otimes \ket{g}_{a_1}\otimes \ket{g}_{a_{1}} \rightarrow \ket{\text{HCS}_{2}^{+}(\alpha)}_{1,2}$.\label{fig:circuit}}
\end{figure}

First, note that one can factorize the CNOT gate on $\mathcal{K} \otimes \mathcal{K}$ into the following product of local Hadamard gates and conditional $\sigma_{z}$ gate: \begin{equation}\text{CNOT} = (\mathbb{I}\otimes H)(\ket{\psi_{+}}\bra{\psi_{+}}\otimes \mathbb{I} + \ket{\psi_{-}}\bra{\psi_{-}}\otimes \sigma_{z})(\mathbb{I}\otimes H).\end{equation} We have already described the procedure for applying a Hadamard gate to the field via the local coupling of the field mode and transmon qubit; hence, we take the initial state to be $H\ket{\psi_{+}}_{1}\otimes H\ket{\psi_{+}}_{2} \otimes \ket{g}_{a_1}\otimes \ket{g}_{a_{1}}$ (where we now explicitly include the field mode labels $1$, $2$ and the transmon qubit mode labels $a_{1}$, $a_{2}$) and show how to implement the conditional $\sigma_{z}$ gate. Let an orthonormal basis for a transmon qubit Hilbert space be taken as $\lbrace \ket{g}, \ket{e}\rbrace$. A quantum circuit diagram showing our method for indirectly performing the CNOT gate on the initial product state is shown in Fig. \ref{fig:circuit}. 
In this circuit, the first field/qubit operation is a $\pi$ rotation of qubit $a_{1}$ conditioned on the parity of field mode 1 and is labeled in Fig. \ref{fig:circuit} by the operation with the $P$ superscript. Explicitly, this unitary is given by
\begin{equation}
\ket{\psi_{+}}_{1}\bra{\psi_{+}}_{1} \otimes \mathbb{I} + \ket{\psi_{-}}_{1}\bra{\psi_{-}}_{1} \otimes \sigma_{x}.
\end{equation}
A similar conditional transformation has been achieved experimentally in Ref.\cite{schoelkopfcat}. This transformation should be followed by a CNOT gate between the qubit modes $a_{1}$ and $a_{2}$, as shown; we assume that this gate is accessible with high fidelity by precise control of the qubit-qubit state. At this point, the full normalized state is given by
\begin{eqnarray}
&{}&{1\over 2}\left( (\ket{\psi_{+}}_{1}\otimes \ket{\psi_{+}}_{2} + \ket{\psi_{+}}_{1}\otimes \ket{\psi_{-}}_{2})\otimes \ket{g}_{a_{1}}\otimes \ket{g}_{a_{2}} \nonumber \right. \\ &+& \left. (\ket{\psi_{-}}_{1}\otimes \ket{\psi_{+}}_{2} + \ket{\psi_{-}}_{1}\otimes \ket{\psi_{-}}_{2})\otimes \ket{e}_{a_{1}}\otimes \ket{e}_{a_{2}} \right).
\end{eqnarray}
The next step is a $\pi$ rotation of field 2 conditioned on the state of qubit $a_{2}$. This operation has been implemented in the experiment reported in Ref.\cite{schoelkopfcat}. Recall from Section \ref{sec:metrouse} that the $\pi$ phase rotation operator acts like $\sigma_{z}$ in the subspace $\mathcal{K}$. Applying again the CNOT gate between qubits $a_{1}$ and $a_{2}$, followed by the parity-conditioned $\pi$ qubit rotation, gives the desired CNOT gate in $\mathcal{K}\otimes \mathcal{K}$. Finally, applying the local Hadamard operator $\mathbb{I}\otimes H$ produces the output state $\ket{\text{HCS}_{2}^{+}(\alpha)} \otimes \ket{g}_{a_{1}}\otimes \ket{g}_{a_{2}}$.

The above method for generating $\ket{\text{HCS}_{2}^{+}(\alpha)}$ is not the most efficient possible. It would be favorable to utilize a single qubit or few-level mode which can be sequentially entangled with both fields \cite{su}.

\section{More exotic hierarchical superpositions}
The notion of hierarchical cat states can be extended to deeper levels of hierarchy. The principal motivation for an analysis of these states comes from the theory of quantum error correction, which makes use of encoded states to strengthen quantum information against unwanted decoherence.  In Refs. \cite{dur1,dur2}, a class of ``concatenated'' GHZ states of the form $\ket{\text{C-GHZ}^{\pm}_{M,N}}:= 1/\sqrt{2}(\ket{\text{GHZ}_{N}^{+}}^{\otimes M} \pm \ket{\text{GHZ}_{N}^{-}}^{\otimes M} )$ were introduced as entangled states which are relatively stable to local noise compared to the full GHZ state $\ket{\text{GHZ}_{NM}^{+}}$. An analog of the C-GHZ states of $(\mathbb{C}^{2})^{\otimes MN}$ can be constructed from HCS in $\ell^{2}(\mathbb{C})^{\otimes MN}$ by forming: \begin{equation}
\ket{\text{C-HCS}^{\pm}_{M,N}} := {1\over \sqrt{2}}\left( \ket{\text{HCS}_{N}^{+}}^{\otimes M} \pm \ket{\text{HCS}_{N}^{-}}^{\otimes M} \right).
\end{equation}
This state can retain coherence on the scale of $N$ modes even after global coherence on the scale of all $MN$ modes has been lost. The C-HCS states are expected to be useful as encoded photonic states for continuous variable quantum error correction schemes. Of course, the entangled coherent states can be concatenated in a similar way:
\begin{equation}
\ket{\text{C-ECS}_{M,N}^{\pm}(\alpha)}:={1\over \sqrt{2}}\left( \ket{\text{ECS}_{N}^{+}(\alpha)}^{\otimes M} \pm \ket{\text{ECS}_{N}^{-}(\alpha)}^{\otimes M} \right).
\end{equation}

It also follows from the basic theory of quantum binary distinguishability that the optimal projection-valued measurement for distinguishing $\ket{\alpha}^{\otimes N}$ from $\ket{-\alpha}^{\otimes N}$ has elements $\lbrace \ket{\text{C-ECS}_{1,N}^{+}(\alpha)}\bra{\text{C-ECS}_{1,N}^{+}(\alpha)} , \ket{\text{C-ECS}_{1,N}^{-}(\alpha)}\bra{\text{C-ECS}_{1,N}^{-}(\alpha)}\rbrace$. The C-HCS and C-ECS states are robust quantum resources in the sense that if coherence is lost among the $M$ blocks of $N$ single-mode systems, a statistical mixture of $N$-mode entangled states remains. To lose all entanglement, the intermode coherence in $\mathcal{H}^{\otimes N}$ must subsequently be lost. In a higher-order hierarchical cat state, these ``shells'' of coherence degrade according to the strengths of local and nonlocal interactions. It has been suggested to generate $\ket{\text{C-GHZ}^{+}_{M,N}}$ in spin-1/2 chains by application of the 2-local M\o lmer-Sorensen unitary gate to the $NM$-mode GHZ state $1/\sqrt{2}( \ket{0}^{\otimes MN} + \ket{1}^{\otimes MN})$  \cite{dur2}. Efficient preparation of hierarchically encoded entangled states of $(\ell^{2}(\mathbb{C}))^{\otimes NM}$ represents a great challenge for continuous variable quantum information processing.

The author acknowledges support from the Howard W. Crandall Memorial Fund at UC Berkeley.

\bibliography{hierarchrefs.bib}

\end{document}